\documentclass[fleqn,usenatbib]{mnras}

\usepackage{newtxtext,newtxmath}
\usepackage[T1]{fontenc}

\DeclareRobustCommand{\VAN}[3]{#2}
\let\VANthebibliography\thebibliography
\def\thebibliography{\DeclareRobustCommand{\VAN}[3]{##3}\VANthebibliography}

\usepackage{graphicx}	% Including figure files
\usepackage{amsmath}	% Advanced maths commands
\usepackage{newtxmath}

\usepackage{blindtext}
\usepackage[normalem]{ulem}
\usepackage{lipsum}

\newcommand{\rshm}{$R_{\rm SHM}$} % Stellar Half Mass Radius
\newcommand{\radialTimescale}{\tau_{\rm GM}}
\newcommand{\rbreak}{$R_{\rm break}$}

\defcitealias{hemler_gas-phase_2021}{H21} 
\defcitealias{springel_cosmological_2003}{SH03}
\def\citeapos#1#2{\citeauthor{#1} (\citeyear{#1}, #2)}

\newif\ifedits
\editsfalse

\ifedits
    \newcommand{\edit}[1]{{\color{red} #1}}
    \newcommand{\editcolor}{red}
\else
    \newcommand{\edit}[1]{#1}
    \newcommand{\editcolor}{black}
\fi

\title[Metallicity Break Radii]{Gas-phase metallicity break radii of star-forming galaxies in IllustrisTNG}

\author[Garcia et al.]{Alex M. Garcia$^{1}$\thanks{E-mail: alexgarcia@ufl.edu},
Paul Torrey$^{1}$,
Z. S. Hemler$^{2}$,
Lars Hernquist$^{3}$,
Lisa J. Kewley$^{3,4,5}$,\newauthor
Erica J. Nelson$^{6}$, 
Kathryn Grasha$^{4,5}$\thanks{ARC DECRA Fellow},
Henry R. M. Zovaro$^{4}$,
Qian-Hui Chen$^{4,5}$
\\
$^{1}$Department of Astronomy, University of Florida, 211 Bryant Space Sciences Center, Gainesville, FL 32611, USA \\
$^{2}$Department of Astrophysical Sciences, Princeton University, Peyton Hall, Princeton, NJ, 08544, USA \\
$^{3}$Institute for Theory and Computation, Harvard-Smithsonian Center for Astrophysics, Cambridge, MA 02138, USA \\
$^{4}$Research School of Astronomy \& Astrophysics, Australian National University, Canberra, Australia, 2611 \\
$^{5}$ARC Centre of Excellence for All Sky Astrophysics in 3 Dimensions (ASTRO 3D) \\
$^{6}$Department for Astrophysical and Planetary Science, University of Colorado, Boulder, CO 80309, USA
}

\date{Accepted XXX. Received YYY; in original form ZZZ}

\pubyear{2022}

\begin{document}
\label{firstpage}
\pagerange{\pageref{firstpage}--\pageref{lastpage}}
\maketitle

\begin{abstract}
We present radial gas-phase metallicity profiles, gradients, and break radii at redshift $z=$ 0 - 3 from the TNG50-1 star-forming galaxy population.
These metallicity profiles are characterized by an emphasis on identifying the steep inner gradient and flat outer gradient.
From this, the break radius, {\rbreak}, is defined as the region where the transition occurs.
We observe the break radius having a positive trend with mass that weakens with redshift.
When normalized by the stellar half-mass radius, the break radius has a weaker relation with both mass and redshift.
To test if our results are dependent on the resolution or adopted physics of TNG50-1, the same analysis is performed in TNG50-2 and Illustris-1.
We find general agreement between each of the simulations in their qualitative trends; however, the adopted physics between TNG and Illustris differ and therefore the breaks, normalized by galaxy size, deviate by a factor of $\sim$2.
In order to understand where the break comes from, we define two relevant time-scales: an enrichment time-scale and a radial gas mixing time-scale.
We find that \rbreak{} occurs where the gas mixing time-scale is $\sim$10 times as long as the enrichment time-scale in all three simulation runs, with some weak mass and redshift dependence.
This implies that galactic disks can be thought of in two-parts: a star-forming inner disk with a steep gradient and a mixing-dominated outer disk with a flat gradient, with the break radius marking the region of transition between them.

\end{abstract}


%%%%%%%%%%%%%%%%%%%%%%%%%%%%%%%%%%%%%%%%%%%%%%%%%%

%%%%%%%%%%%%%%%%% BODY OF PAPER %%%%%%%%%%%%%%%%%%

\section{Introduction}
\label{sec:intro}

The abundances of metals within galaxies, known as a system's metallicity, provide an essential tool for constraining galaxy formation models.
Since the overwhelming majority of metal production is associated with stars -- either on the main sequence or during their late stage stellar evolution -- overall metal abundances are set by the star-formation history of the system. 
Newly synthesized metals can then be expelled back into the interstellar medium (ISM) in the gas phase \citep[e.g.][]{friedli_influence_1994} through stellar winds and supernovae. 
After being injected back into the ISM, the metals will mix with the galactic gas and can be redistributed throughout the galaxy.
Metals can be ejected from the galactic disc via galactic winds \citep{lacey_chemical_1985,koeppen_evolution_1994}, 
mixed via turbulance \citep[e.g.][]{elmegreen_formation_1999,burkhart_density_2009},
or moved in bulk throughout the galaxy, e.g., as in galaxy mergers \citep{rupke_gas-phase_2010,torrey_metallicity_2012}. 
All of these effects can be classified as two different mechanisms: enrichment and gas redistribution. 
Enrichment raises the metallicity locally and gas redistribution moves those metals throughout a galaxy. 
In this paper, we focus on understanding and delineating the regions within a galaxy where enrichment versus gas mixing dominate.

This link between the evolution of galaxies and their metal contents gives rise to a fundamental relationship: the mass-metallicity relation (MZR). 
The MZR describes a tight (scatter of only $\sim$0.1 dex) correlation of increasing gas-phase metallicity with increasing galaxy stellar mass across several orders of magnitude \citep[][]{lequeux_chemical_1979,tremonti_origin_2004,lee_extending_2006}. 
The MZR appears to follow three different power laws on different mass ranges. In the low-mass regime ($\log[M_*/M_\odot] < 9.5$) there exists a shallow power-law which, at more intermediate-masses, transitions to a steeper relation ($9.5 < \log[M_*/M_\odot] < 10.5$). 
At higher masses ($\log[M_*/M_\odot] > 10.5$), the power-law flattens out significantly almost to a constant value of 12 + log(O/H) $\simeq$ 8.8 \citep[][]{blanc_characteristic_2019}. 
These differing power laws on different mass scales are thought to be indicative of the efficiency of the aforementioned processes of enrichment and mixing scaling with galaxy mass. 
The MZR has been observed in galaxies back to $z \sim 4$ \citep[][even assert trends can be seen as far as $z \simeq 7$]{finkelstein_candels_2012}; with increasing redshift there is an observed decrease in the overall metallicity of galaxies \citep[e.g.][]{savaglio_gemini_2005,erb_mass-metallicity_2006,maiolino_amaze_2008,zahid_mass-metallicity_2011,zahid_universal_2014}. 
This is a feature of higher redshift galaxies having had less time for their stellar populations to evolve and not only synthesize heavier metals, but also eject them back into the ISM. 
% Additionally, not only does the normalization of the MZR change, so too do the power laws. 
% \cite{savaglio_gemini_2005} suggest that as galaxies grow, more massive galaxies reach higher metallicities faster than low-mass galaxies causing the MZR to be flatter at low redshift. 
In all, the MZR offers a fundamental observational constraint on models of galaxy evolution.

While the canonical MZR is an incredibly powerful tool for studying bulk galaxy trends, it simplifies galaxies by assuming an effective, galactic average of the gas-phase metallicity \citep{tremonti_origin_2004}. 
Galaxies are not homogeneous and do not evolve uniformly in space or time. 
Different regions of galaxies have different dominant physical processes that can drive the metal contents much higher or lower than the galaxy average. 
Thus, with further investigation of individual galaxies, radial metallicity gradients have been seen to exist. 
At low redshift, it has been observed that, generally, galaxies have decreasing metallicity with increasing radius \citep[e.g.][]{searle_evidence_1971,dennefeld_supernova_1981,zaritsky_h_1994,magrini_metallicity_2007,fu_origin_2009}. 
This decrease indicates that the gradients of the gas-phase metallicity for individual galaxies are predominantly negative.
This negative gradient is attributed to inside-out galaxy growth, wherein galaxies' stellar populations in the inner-most regions form and evolve before further out regions \citep[e.g.][]{van_den_bosch_formation_1998,prantzos_chemo-spectrophotometric_2000,perez_evolution_2013}.
The more evolved stellar populations at the centre of galaxies have produced more massive stars, which in-turn produce more metals, sooner than the outskirts of the galaxy.

For low redshift systems, this customary explanation of inside-out growth proves a good metric for explaining metallicity gradients.
At higher redshifts, however, the narrative becomes more complicated as some contention exists between theory and observations.
Several recent observational works have found that most galaxies at $z \sim 0.6-3$ display a wide variety of gradients, not just negative \citep[e.g.][]{cresci_gas_2010,queyrel_massiv_2012,swinbank_properties_2012,wuyts_evolution_2016}.
This is in direct contrast to predictions from simulations arguing that gradients of galaxies are negative, and even strengthen, at higher redshift \citep[][though this is dependent on the strength of implemented feedback models, see \citeauthor{gibson_constraining_2013} \citeyear{gibson_constraining_2013}]{pilkington_metallicity_2012,hemler_gas-phase_2021,yates_l-galaxies_2021}.
This contention is still an active area of research and with the next generation of instruments (JWST, ELT, etc), the ability to make spatially resolved high-redshift measurements will substantially increase, allowing for dramatically improved constraints for theoretical models.

Regardless of redshift, metallicity gradients, both in observational studies \citep[e.g.][]{magrini_metallicity_2007,jones_measurement_2010,yuan_metallicity_2011,swinbank_properties_2012,sanchez_characteristic_2014,grasha_metallicity_2022} and simulations \citep[e.g.][]{tissera_oxygen_2019,collacchioni_effect_2020,hemler_gas-phase_2021}, are usually characterized in terms of a single value with a linear least-squares fit.
It is customary, particularly at low redshift, to define a specific region (typically within [0.5,2.0]$R_e$, where $R_e$ is the effective radius of the galaxy, but other conventions exist) to define a metallicity gradient.
At higher redshift, however, this practice is less common as these profiles seem well-fit over the extent of the system, though this could be a feature of decreased spatial resolution. 
Despite this convention of measuring gradients linearly throughout a disc, it has been noted that outside of these regions the metallicity gradient flattens significantly \citep[][Q.-H. Chen et al. in prep]{martin_oxygen_1995,twarog_revised_1997,carney_elemental_2005,yong_elemental_2005,yong_elemental_2006,bresolin_flat_2009,vlajic_abundance_2009,vlajic_structure_2011,sanchez_characteristic_2014,grasha_metallicity_2022}. 
While these outer flat regions have been seen to exist, most studies up to this point have done little to examine them.

Characterization of the inner regions of the galaxy typically relies upon emission from \ion{H}{II} regions \citep[e.g.][]{shaver_galactic_1983,vilchez_chemical_1988,esteban_keck_2009,grasha_metallicity_2022} and planetary nebulae \citep[e.g.][]{maciel_abundance_1994,maciel_estimate_2003} from a number of different ions, including, but not limited to, [\ion{O}{II}], [\ion{O}{III}], [\ion{S}{II}], and [\ion{N}{II}] \citep[e.g.][]{shaver_galactic_1983,perez-montero_impact_2009,peimbert_nebular_2017}. Measurements of metallicity from emission lines are often broken into two different categories: direct method and strong line method \citep[advantages and disadvantages covered in depth in ][]{kewley_understanding_2019}.
The direct method \citep[outlined fully in][]{perez-montero_ionized_2017} uses the fluxes of auroral lines to determine electron temperatures.
Uncertainty in the metallicity measurements is dominated primarily by uncertainty in the electron temperature. Moreover, auroral emission lines are intrinsically weak in high metallicity systems \citep[e.g.][]{hoyos_impact_2006}.
In these scenarios where auroral lines are unavailable, the strong line method is the only method of measuring these metallicities.
Even though strong line diagnostics are easier to obtain, the direct method is generally preferred due to systematic biases associated with them \citep[e.g.][]{perez-montero_comparative_2005,kewley_metallicity_2008,stasinska_nebular_2010}.
Additionally, the emission lines used in strong-line methods predominantly trace high-density, star-forming regions of a galaxy, which is the typical window where gradients are defined. 

The highest density star-forming regions are only part of the full picture of galactic evolution, however.
For example, the baryon cycle describes interactions between the star-forming ISM gas, gas within the circumgalactic medium (CGM), and diffuse gas in the intergalactic medium (IGM).
In this cycle, IGM gas cools onto the CGM and then onto the ISM.
Once accreted onto the ISM, the gas provides fuel for star-formation within the disc. 
These stars, in turn, enrich the gas and launch it back into the ISM and CGM via feedback.
Depending on the strength of the feedback, some of this gas escapes into the IGM \citep[e.g.][]{heckman_absorption-line_2000} while more is then effectively `recycled' as it cools back into the ISM \citep[e.g.][]{shapiro_consequences_1976}.
Once back in the ISM, the cycle continues as the enriched gas mixes with new gas accreted onto the disc from the IGM. 

Baryons in the lower-density gas accreting from the CGM and IGM further into the outskirts of the galaxy are difficult to detect using emission line diagnostics \citep[e.g.][]{wijers_abundance_2019,augustin_emission_2019}. Therefore, in order to obtain spatially extended metallicity profiles, other methods of measuring metal contents must be employed.
One method is via absorption line spectroscopy from bright background sources, such as quasars \citep[e.g.][]{werk_cos-halos_2012,werk_cos-halos_2014}. 
For this method to work, background sources (often quasars) need to be coincident with the line of sight of a galaxy which means that such systems are rare. 
Due to their rarity, absorption spectra can stacked into profiles of multiple absorption measurements \citep[][]{norris_oxygen_1983,ellison_enrichment_2000}. 
Absorption spectra that travel through these low-density ($\log N_{\ion{H}{I}} > 20.3 ~{\rm cm}^{-2}$) neutral \ion{H}{I} regions are called damped Lyman-$\alpha$ systems (DLAs) \citep[see][]{wolfe_damped_2005}. 
From the DLAs, the metal lines can be used to determine the metallicity of the cloud. 
Some useful features of these absorption features are that they are independent of redshift, the metallicity of surrounding gas, and excitation state \citep[][all of which are important factors in emission line diagnostics]{peroux_cosmic_2020}. Though absorption features measure the neutral gas metallicities, compared to ionized gas metallicities estimated via emission lines, studies have shown that the two methods have relatively good agreement \citep[][]{christensen_verifying_2014,rahmani_study_2016}.
% Therefore, we explore the possibility of augmenting emission line metallicities with absorption lines to create metallicity profiles that extend deep in the disc of the galaxy.
 
\edit{
These extended metallicity profiles have the potential to help constrain the aforementioned gradient tension at high redshift.
The observed flat gradients at higher redshifts (if they are indeed real and not an artifact of uncertain diagnostics and poor resolution) could be driven by either bulk gas flows within the disc, or by episodic bursty feedback events.
Large radial flows smoothly transport material throughout the galaxy, resulting in a flattening of the overall gradient.
Bursty feedback, on the other hand, is driven by quick bursts of high star formation followed by high gas outflow rates achieving the same effect \citep[see][]{ma_why_2017}.
A differentiating factor between the two models would be where their extended profiles flatten out, which we call the break radius of the system.
Compared to bulk flows, bursty feedback would very quickly radially transport material throughout the entire disc.
This quick homogenisation of the disc would, catastrophically and rapidly, erase the metallicity gradient in the immediate aftermath of strong burst events.
Bulk gas flows, on the other hand, would act more gently and allow for a persistent gradient -- albeit with a shallower slope -- to be present for longer periods of time.
Moreover, as we explore in this paper, the addition of a significant radial gas transport mechanism from turbulence would likely result in the break radius moving to smaller radii which would not necessarily be the case for bursty feedback.
Therefore, spatially distributed metal contents of galaxies, specifically where they flatten out (i.e., the break radius), in non-bursty feedback models (like that in IllustrisTNG) are critical in providing a potentially observational constraint on the extent to which gradient flattening depends on turbulence, but not burstiness in galaxy feedback models.
}

In this paper, we study the radial gas-phase metallicity profiles and gradients from the Illustris and IllustrisTNG cosmological simulations. 
Specifically, we characterize the relationship between the radius that captures the transition of the gradients from steep to flat in terms of traditional measures of galactic size.
We examine this relationship across a large range of galaxy masses as well as study its redshift evolution.

The structure of this paper is as follows. 
In Section \ref{sec:Methods}, we outline the Illustris and IllustrisTNG simulation suites, our method for selecting galaxies, the techniques employed in examining the profiles and gradients of the selected galaxies, and define relevant time-scales. In Section \ref{sec:Results}, we present our findings on the mass and redshift evolution of the break radius, as well as compare the enrichment time-scales of the gas to dynamic radial evolution of the systems. In Section \ref{sec:Discussion}, we offer insight into the effects that our simulation resolution and physics play in our results, discuss how this work relates to the baryon cycle, and speculate on how the trends could be measured observationally. In Section \ref{sec:Conclusion}, we present a summary and state our conclusions.

\begin{figure*}
    \centering
    \includegraphics[width=\linewidth]{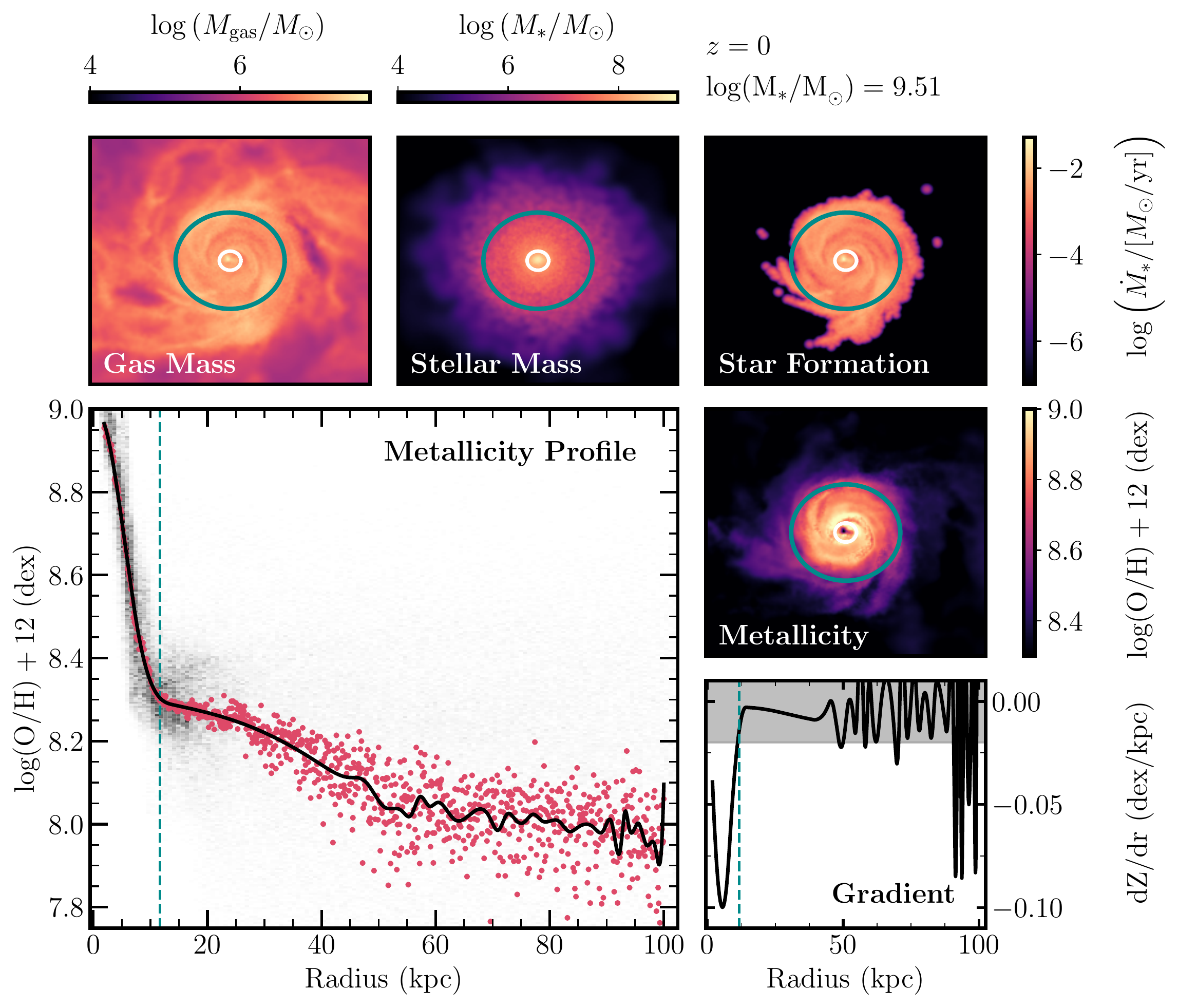}
    \caption{Maps and metallicity profiles for an individual intermediate-mass star-forming galaxy from our IllustrisTNG sample. This individual galaxy is presented as a demonstration of our methods, which are applied to stacked median profiles (opposed to the individual galaxy shown; see Section \ref{subsec:ProfsGrads}). Each of the following four maps extends 30 kpc in either direction from the center of the galaxy. {\it Top Left}: Map of the galaxy gas mass, {\it Top Centre}: map of stellar mass, {\it Top Right}: map of star formation, and {\it Middle Right}: map of gas-phase metallicity. The two circles overplotted on each of the aforementioned four maps correspond to the stellar half-mass radius, {\rshm}, (white) and the identified break radius, {\rbreak}, (green). {\it Bottom Right}: The metallicity gradient, numerically computed from our galaxy profile fitting methodology (see Section \ref{subsec:Fitting}). The gray shaded region represents the threshold value that we designate as the value at which the profile `breaks' (computed using Eqn. \ref{eqn:TNGgrad}), \edit{shown in the profile by the green dashed line}. {\it Bottom Left}: The one-dimensional metallicity profile as a function of galactocentric radius, where the red dots represent the median metallicity at any radius and the black line is a Savitzky-Golay smoothed spline-fit (see Section \ref{subsec:Fitting}, \edit{we note that this is an individual profile and is treated slightly differently than stacked profiles} -- see Appendix \ref{appendix:idvGalaxies}) to the median profile. The green vertical line is the break radius. }
    \label{fig:OooohPretty}
\end{figure*}

\section{Methods}
\label{sec:Methods}

In this paper, we analyse gas-phase metallicity profiles of Illustris and IllustrisTNG galaxies in order to determine the radius at which their gradients transition from steep to shallow.
In this section, we overview the Illustris and IllustrisTNG simulation suites, the selection method of our galaxies, how we generate metallicity profiles and gradients, our fitting methodology of the profiles, and define the relevant time-scales for metallicity evolution within the systems.
Many of the methods employed in this paper follow closely with \citeapos{hemler_gas-phase_2021}{henceforth \citetalias{hemler_gas-phase_2021}}. All measurements have physical units (the exception to this being the box sizes of simulations, which are in comoving units).
As alluded to later in this section, several of the methods are not indicative of a mock observational analysis of Illustris, nor IllustrisTNG, galaxies and thus we do not present the results as such.
Instead, this work aims to provide a theoretical baseline for determining the metallicity break radii of galaxies. 

\subsection{Simulation Details}
\label{subsec:Illustris}

For our analysis, we work in both the Illustris \citep[][]{vogelsberger_model_2013,vogelsberger_introducing_2014,vogelsberger_properties_2014,genel_introducing_2014,torrey_model_2014} and IllustrisTNG \citep[][hereafter TNG]{marinacci_first_2018,naiman_first_2018,nelson_first_2018,pillepich_first_2018,springel_first_2018} cosmological simulation suites, though the bulk of the analysis is in TNG.
Both Illustris and TNG run on moving-mesh hydrodynamical code \textsc{arepo} \citep[][]{springel_e_2010}.
TNG offers an update in the physical models as well as alleviates some deficiencies in the original Illustris simulations.
We employ both the Illustris and TNG simulations in this work as similar, but appreciably different, physical models to identify the physics responsible for setting the location of the metallicity break radius.

The Illustris framework, on which both suites are built, models several important astrophysical processes; namely, star-formation, stellar evolution, chemical enrichment, primordial and metal-line gas cooling, stellar feedback-driven galactic outflows, and supermassive black hole formation, growth, and feedback.
Owing to the limited spatial resolution of large-box cosmological simulations \edit{(see Table~\ref{tab:resolutionTable} for mass resolutions and spacial softening lengths)}, both Illustris and TNG follow the \cite{springel_cosmological_2003} effective equation of state to model the dense, star-forming ISM.
In these regions (set with a threshold density $n_{\rm H} > 0.13 ~{\rm cm}^{-3}$) star-particles are formed following the \cite{chabrier_galactic_2003} initial mass function (IMF) while adopting their metallicity from the ISM from which they form.
As time progresses and stars move off the main sequence, both the mass and metals of the star are injected back into the surrounding ISM.
The stellar lifetime models \citep[adopted from][]{portinari_galactic_1998} depend on the mass and metallicity of the star.
The majority of mass and metals are returned via lower mass stars on the asymptotic giant branch (AGB) ejecting primarily from winds and through massive stars undergoing Type II supernovae (SNe).
% Though Type I supernovae are subdominant in mass ejection, they produce a substantial amount of iron.
Both suites explicitly track the evolution of nine chemical elements (H, He, C, N, O, Ne, Mg, Si, and Fe), but TNG adds a tenth ``other metals'' %{\red Europium?}
item to represent additional metals not explicitly tracked.

\begin{table}
    \centering
    \color{\editcolor}
    \begin{tabular}{lllll}
        \hline
         &                                       & TNG50-1 & TNG50-2 & Illustris-1 \\\hline\hline
        $m_{\rm baryons}$    & $[10^5 M_\odot]$  & $0.85$  & $6.8$   & $12.6$ \\
        $m_{\rm DM}$         & $[10^5 M_\odot]$  & $4.5$   & $36.3$  & $62.6$ \\
        $\epsilon_{\rm *}$   & [pc]              & $288$   & $576$   & $710$  \\
        $\epsilon_{\rm gas}$ & [pc]              & $72$    & $147$   & $710$  \\
        $\epsilon_{\rm DM}$  & [pc]              & $288$   & $576$   & $1420$ \\\hline
    \end{tabular}
    \caption{Mass resolution and maximum physical spatial softening for Illustris TNG50-1, -2, and Illustris-1 for baryons and dark matter. Values for Illustris from \citeauthor{vogelsberger_introducing_2014} \citeyear{vogelsberger_introducing_2014} Table 1. Values for TNG from \citeauthor{pillepich_first_2019} \citeyear{pillepich_first_2019} Table B1, see those works for more details.}
    \label{tab:resolutionTable}
\end{table}

The differences between Illustris' and TNG's physical models and methods are well documented \citep[see][for a complete reference]{weinberger_simulating_2017,pillepich_simulating_2018}. We therefore offer only an overview of the differences for this analysis.
For our purposes, the important differing physical implementations are galactic winds and stellar evolution.
These processes are vital in setting and evolving the metallicity of a system. The galactic winds in TNG are isotropic, being emitted with no preferred direction, whereas Illustris implements a bipolar model with a wind preferentially along the rotation axis.
Additionally, the wind speeds in TNG are set %by the local dark matter velocity dispersion as well as 
with an explicit redshift dependency.
This redshift scaling increases star-formation suppression by winds in low redshift systems compared to Illustris.
An additional wind velocity floor is added in TNG having the effect of making stellar feedback more important at high redshifts.
In terms of the stellar evolution, one large change is that the minimum mass for a core-collapse supernova is modified from 6$M_\odot$ in Illustris to 8$M_\odot$ in TNG.
With both assuming a Chabrier IMF, this change in threshold leads to an approximately 30\% \edit{decrease} in Type II SNe in TNG.
This increase in SNe for TNG has a direct effect on the metallicity as AGB stars form the vast majority of s-process elements, whereas core-collapse SNe produce r-process elements.
Thus, any change in the amount of SNe in the model has a direct impact on the chemical abundances of the ISM.

\subsubsection*{Illustris}

The original Illustris suite is comprised of three cosmological simulations with a box size of (106.5 Mpc)$^3$.
The three different runs of Illustris vary in the number of resolution elements.
To distinguish each resolution variation a number is added to the end of the simulation name, with the higher numbers corresponding to the lower resolution runs. Illustris-1, the highest resolution run, has $3\times1820^3$ resolution elements, while the other two runs (Illustris-2 and -3) have $3\times910^3$ and $3\times455^3$, respectively.
For the purposes of this paper, we will be using Illustris-1 as a measure of the influence of the adopted physics compared to the TNG simulation. 

\subsubsection*{IllustrisTNG}

TNG (The Next Generation) is the follow-up to the original Illustris suite.
Whereas the Illustris runs comprise one box size, TNG includes runs of three different box sizes, each with its resolution variations.
The names of these simulations roughly correlate to their box sizes in Mpc -- TNG50 (51.7 Mpc)$^3$, TNG100 (110.7 Mpc)$^3$, and TNG300 (302.6 Mpc)$^3$. The same naming convention from Illustris for the resolution variations is used in TNG.
TNG50-1, the highest resolution run, has $2\times 2160^3$ resolution elements and TNG50-2 $1080^3$ elements. The larger boxes, TNG100 and TNG300, have $2\times 1820^3$ and $2 \times 2500^3$ resolution elements in their highest resolution runs, respectively.
In this work, we analyse data from the smallest box-size simulations, TNG50, in particular the two highest-resolution, TNG50-1 and TNG50-2
(For details about TNG50, see \citeauthor{pillepich_first_2019} \citeyear{pillepich_first_2019}; \citealt{nelson_illustristng_2019,nelson_first_2019}).
% ,nelson_first_2019}.)
We utilize the two simulations with the same box size as a study of the resolution effects on the results.

\subsection{Galaxy Selection}
\label{subsec:GalSelect}

Gravitationally-bound substructures are identified in both Illustris and TNG using the \textsc{subfind} algorithm \citep[][]{springel_hydrodynamic_2001,dolag_substructures_2009}, which relies on the Friends-of-Friends (FoF) method \citep[][]{davis_evolution_1985} to find parent groups.
For this analysis, we impose stellar mass limits of $8.5 < \log(M_*/M_\odot) < 10.9$ on our galaxies.
We break up our galaxies into mass bins of width 0.5 dex with step size of 0.1 dex in order to continuously examine the relation with respect to mass (see Table \ref{tab:Simulated Distances}).
We impose the lower mass limit to allow enough resolution elements to define a gradient.
To this end, we find that $\gtrsim 10^4$ resolution elements are needed to sufficiently determine the gradient of the galaxies.
Additionally, we set a minimum gas mass threshold of galaxies to $\log(M_{\rm gas}/M_\odot) > 8.5$.

Following \cite{donnari_star_2019}, \cite{pillepich_first_2019}, \cite{nelson_spatially_2021}, and \citetalias{hemler_gas-phase_2021}, we define a specific star formation main sequence (sSFMS) with integrated specific star formation rates (sSFR).
The sSFMS is characterized with a median sSFR relation for galaxies $\log (M_*/M_\odot) \leq {10.2}$ in mass bins of width 0.2 dex.
Above this mass threshold, we extrapolate a linear-least squares fit through these lower mass medians.
We then define a galaxy as star-forming if its sSFR is above those on the sSFMS or less than 0.5 dex below it at its given mass.
The non-star-forming galaxies are excluded from our sample.
This allows a more fair comparison for observations since metal abundances are most easily measured from star-forming \ion{H}{II} regions in the inner portions of galaxies \citep[][]{kewley_understanding_2019}.
Though we propose the use of absorption features later in this work, they are not sufficient on their own to generate extended metallicity profiles.
Additionally, we restrict our sample to only central galaxies, excluding satellite galaxies.
Based on these prescriptions, for TNG50-1, we have a sample of 2,751 galaxies at $z=0$ and 1,627 galaxies at $z=3$.
\edit{Maps of the gas mass, stellar mass, star formation, and gas-phase metallicity of a galaxy from our sample are shown in Figure \ref{fig:OooohPretty}.}

\begin{figure}
    \includegraphics[width=\linewidth]{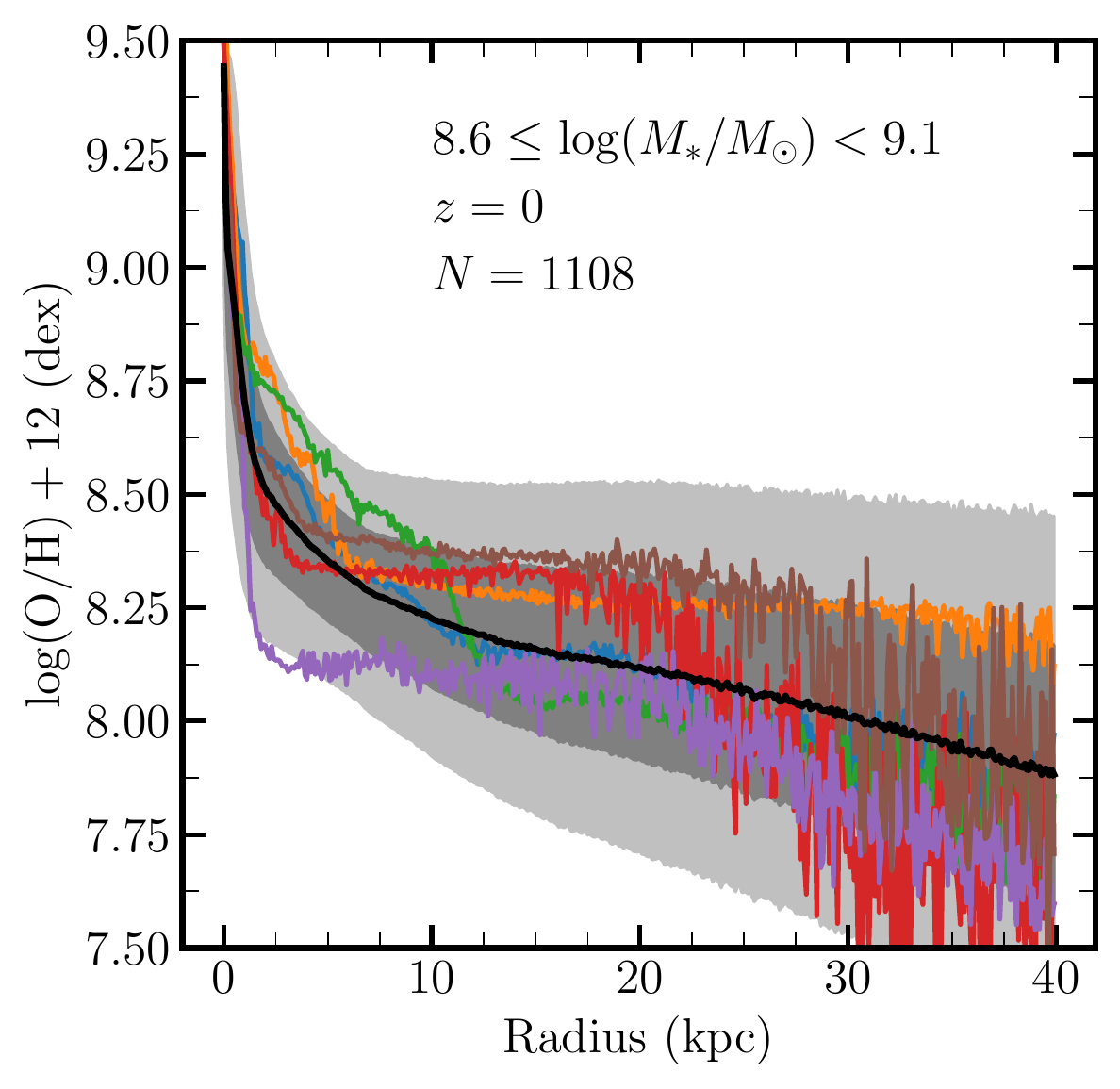}
    \caption{Example stacked median metallicity profile of the galaxies in the stellar mass $8.6 \leq \log (M_*/M_\odot) < 9.1$ mass bin at redshift $z=0$ as a function of galactic radius. The colored lines represent individual median profiles that are constituents of this mass bin, while the black line represents the stacked median profile, the median of all the individual median profiles. The shaded regions represent 1$\sigma$ and 2$\sigma$ deviations from the median. These profiles extend out to 40.0 kpc; see Table \ref{tab:Simulated Distances} for the extent of each mass bin's profile.}
    \label{fig:median_example}
\end{figure}

\subsection{Characteristic Radii}
\label{subsec:Radii}

In our analysis, we define several characteristic radii with respect to the centre of each galaxy.
The centre of a galaxy is defined by the location of the potential minimum.
The first characteristic radius follows from \citetalias{hemler_gas-phase_2021} (referred to as $R_{\rm in}$ in that work): the 3D radius encapsulating the inner-most 10\% of the star-formation rate of each galaxy, for which we adopt the name $R_{\rm SFR,10}$.
We utilize $R_{\rm SFR,10}$ to begin our examination of the profiles as some, particularly high-mass, galaxies have significantly hollowed out inner regions due to AGN feedback \citep{nelson_spatially_2021}.
These inner regions dominated by AGN feedback are relatively devoid of gas, therefore the metallicity profiles in these regions cannot be robustly measured and are omitted from our analysis.
Another characteristic radius that we utilize is the stellar half-mass radius, henceforth $R_{\rm SHM}$, defined as the radius enclosing 50\% of the stellar mass in the galaxy.
In TNG, the stellar half-mass radii have been calculated and made available by \cite{genel_size_2018}; details on how the half-mass radius is calculated can be found within that work.
A number of different observational studies have shown that metallicity gradients, when normalized by the effective radius of the galaxy, are fairly similar 
\citep[e.g.][]{sanchez_integral_2012,sanchez_characteristic_2014,sanchez-menguiano_shape_2016}.
In simulations, however, it has been suggested that gradients scale with a different quantity, $R_{\rm SFR}$ \citepalias[][discussed further in Section \ref{subsec:Fitting}]{hemler_gas-phase_2021}.
% (e.g. \citeauthor{sanchez_integral_2012} \citeyear{sanchez_integral_2012}, \citeyear{sanchez_characteristic_2014}; \citeauthor{sanchez-menguiano_shape_2016} \citeyear{sanchez-menguiano_shape_2016}).
As we detail later in Section \ref{subsec:MassAndRedshiftEvo}, we find that galaxy break radii scale in a similar fashion with {\rshm}.

Since part of this analysis involves stacking individual metallicity profiles (see Section \ref{subsec:ProfsGrads}), we set standardized endpoints for galaxies within a certain mass bin, designated $r_{\rm max}$ (see Table \ref{tab:Simulated Distances}).
This choice is primarily for convenience in stacking, but after varying $r_{\rm max}$ we find that, as long as the profiles extend past the gradient transition region, the selection does not significantly impact our results\footnote{The only galaxies that do not adhere to this convention are the Illustris-1 $z=0$ galaxies.
When using this prescription on those galaxies the transition region was not captured, therefore all of those galaxies are simulated out to 100 kpc}.
The chosen values are drawn primarily from the TNG50-1 $z=0$ galaxies, as redshift increases the necessary length for galaxies to capture this gradient transition decreases, but we adopt the $z=0$ convention for higher redshifts for simplicity.

\subsection{Galaxy Orientation}
\label{subsec:Incline}

Following from \citetalias{hemler_gas-phase_2021} and \cite{ma_why_2017}, a key component of this analysis is orienting all of the galaxies to the face-on orientation.
In order to rotate the galaxies, we first find the inclination angle of the galaxy. This angle is defined with respect to a vector normal to the angular momentum vector of the galaxy.
The angular momentum vector is computed from the sum of all the angular momenta of the star-forming gas (density greater than TNG star-formation density threshold) within the disk.
The inclination angle is used to rotate the galaxy into the face-on orientation for analysis.
This face-on orientation allows us to look at metallicity, and thus the metallicity gradient, as having a strictly radial dependency \edit{(see Figure~\ref{fig:OooohPretty} for face-on galaxy orientation as well as a purely radial metallicity profile and gradient)}.
A number of studies have shown that galaxies have been seen to have vertical gradients as well \citep[e.g.][]{marsakov_formation_2005,marsakov_formation_2006,soubiran_vertical_2008,pilkington_metallicity_2012}.
By aligning the galaxies to the face-on orientation, we lose this information, therefore for the purposes of this analysis we will examine only the radial component of metallicity gradients.

\subsection{Median Metallicity Profiles and Gradients}
\label{subsec:ProfsGrads}

\begin{figure}
    \centering
    \includegraphics[width=\linewidth]{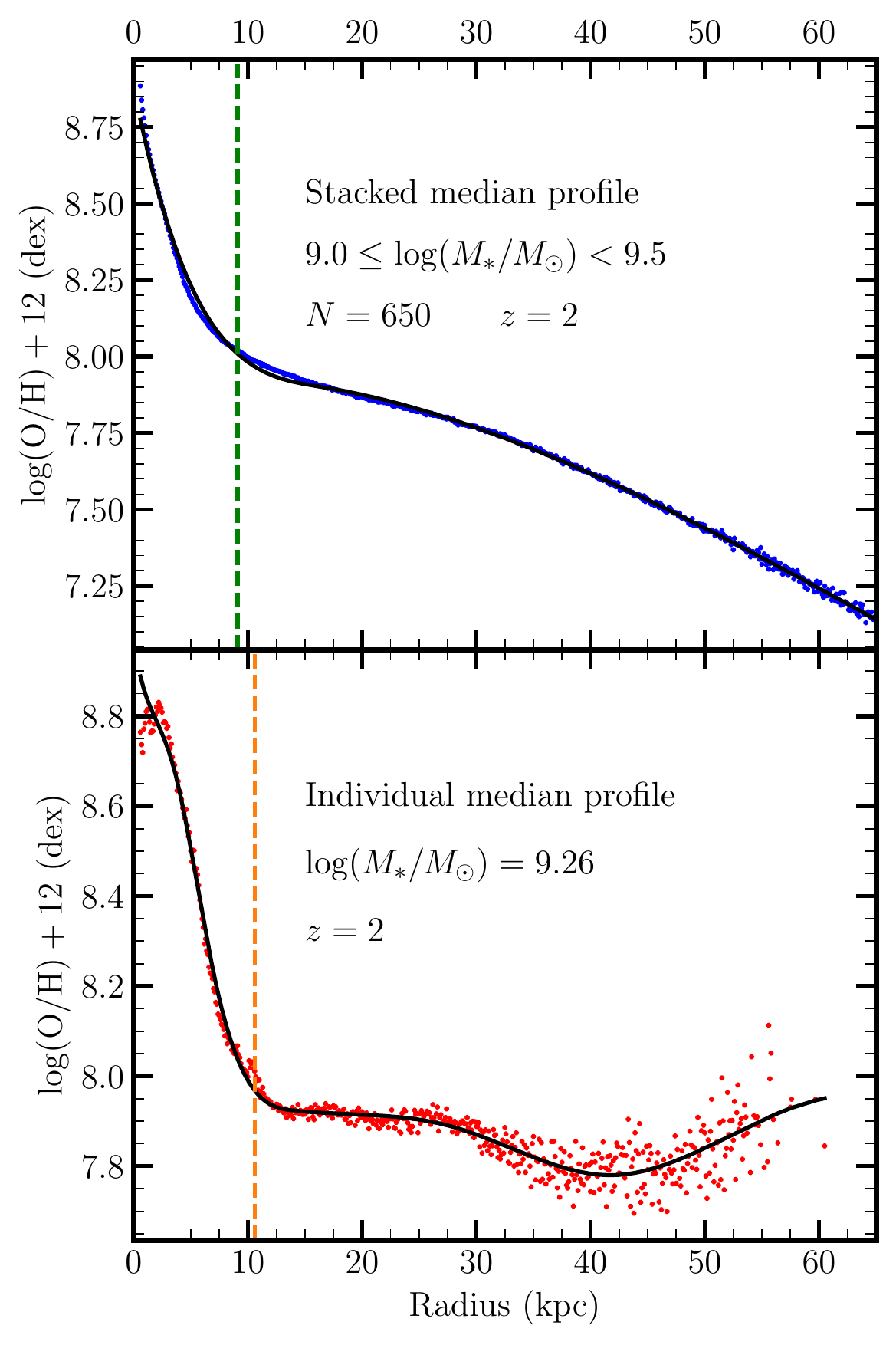}
    \caption{Our method of fitting gas-phase metallicity profiles as a function of galactocentric radius. {\it Top}: A stacked median profile (the $9.0 \leq \log(M_*/M_\odot) < 9.5$ mass bin at $z=2$). The blue points represent the profile generated by stacking and taking the median of the individual profiles within a given mass bin. The black line is our spline fit to the data and the dashed line is the break \edit{(both defined in section \ref{subsec:Fitting})}. {\it Bottom}: Our method applied to an individual galaxy from the same mass bin. Similar to the top panel, the red points represent the median profile, the black line is our spline fit (note that the fitting methodology is slightly different for individual profiles: see Appendix \ref{appendix:idvGalaxies}), and the dashed orange line is the identified break. Note that this work primarily focuses on the analysis of stacked galaxy profiles and that this individual galaxy is shown just for demonstration purposes. We examine individual galaxy profiles in Appendix \ref{appendix:idvGalaxies}.}
    \label{fig:median_profiles}
\end{figure}

\begin{table}
    \centering
    \begin{tabular}{cccc}
        \hline
        $M_{*}^{\rm min}$ & $M_{*}^{\rm max}$ & $r_{\rm max}$ & $N$ \\
        $\log [\frac{M_{*}}{M_\odot}]$ & $\log [\frac{M_{*}}{M_\odot}]$ & (kpc) & TNG50-1 \\\hline\hline
        8.5 & 9.0 & 40.0 & 1156\\
        8.6 & 9.1 & 40.0 & 1108\\
        8.7 & 9.2 & 40.0 & 1072\\
        8.8 & 9.3 & 40.0 & 987\\
        8.9 & 9.4 & 40.0 & 902\\
        9.0 & 9.5 & 65.0 & 806\\
        9.1 & 9.6 & 65.0 & 703\\
        9.2 & 9.7 & 65.0 & 600\\
        9.3 & 9.8 & 65.0 & 552\\
        9.4 & 9.9 & 65.0 & 482\\
        9.5 & 10.0 & 100.0 & 423\\
        9.6 & 10.1 & 100.0 & 384\\
        9.7 & 10.2 & 100.0 & 347\\
        9.8 & 10.3 & 100.0 & 300\\
        9.9 & 10.4 & 100.0 & 289\\
        10.0 & 10.5 & 150.0 & 263\\
        10.1 & 10.6 & 150.0 & 233\\
        10.2 & 10.7 & 150.0 & 205\\
        10.3 & 10.8 & 150.0 & 176\\
        10.4 & 10.9 & 150.0 & 131\\\hline
    \end{tabular}
    \caption{Each of the mass bins, their radial extent $r_{\rm max}$, and the total number of galaxies, $N$, in each bin for the TNG50-1 sample at $z = 0$. The overlap in the bins corresponding to multiple $r_{\rm max}$ values is a feature of the selection method of the sample. We select all profiles within the mass range, obtain the data, increase the mass limits and obtain the data with our new constraints. Note that the number of galaxies in each bin in TNG50-2 is very similar to the number in TNG50-1, since they are the same box size. However, the number of galaxies in each bin is approximately an order of magnitude larger owing to Illustris' larger box size.}
    \label{tab:Simulated Distances}
\end{table}

Using shells of radius 0.1 kpc and setting the upper radial bound, $r_{\rm max}$ (see Table \ref{tab:Simulated Distances} and Section \ref{subsec:Radii}), we generate a median metallicity profile as a function of galactic radius for each galaxy in our sample.
To ensure that each shell has a statistically robust number of particles, we set a minimum number within a given region to 6 particles.
The gas-phase metallicity value at each of these radial bins is defined by its oxygen-to-hydrogen ratio, a quantity that is readily available to us in both Illustris and TNG.
We follow the convention of defining the relative abundance ratio, $\epsilon$, of species $X$ and hydrogen to be

\begin{equation}
    \epsilon(N_X,N_H) = 12 + \log(N_X/N_H),
\end{equation}

\noindent where the number of element $X$ and Hydrogen nuclei are defined by $N_X$ and $N_H$, respectively.

% While \citetalias{hemler_gas-phase_2021} restricts their analysis to only star-forming regions (set by an ISM density greater than 0.13 cm$^{-3}$), we do not impose this limit.
% As previously mentioned in Section \ref{sec:intro}, this work suggests a possible connection with nebular emission and absorption spectra to create extended metallicity profiles.
% Therefore, we do not restrict our analysis to star-forming regions since, theoretically, all information about the metallicity should be available for examination within a galaxy.

As a tool to study general population trends of the break radius, we create stacked median profiles.
Stacking galaxy information by taking a median or mean is a fairly common practice for understanding bulk properties of galaxies observationally \citep[e.g.][]{andrews_mass-metallicity_2013,zahid_stellar_2017} and theoretically \citepalias[e.g.][]{hemler_gas-phase_2021}.
We first create individual median metallicity profiles for each individual galaxy, as described above.
Once the individual median profiles are obtained, we further combine all the individual profiles in a given mass bin to create a stacked profile.
Stacked profiles are constructed by taking the median of all the individual median profiles in their shells of 0.1 kpc \edit{(we note that this is slightly different than stacking done in observational studies; see Appendix~\ref{appendix:stacking})}. 
This gives us one profile that roughly characterizes all of the galaxies for each mass bin. 
Henceforth this will be referred to as the stacked median profile. See Figure \ref{fig:median_example} for a schematic of the stacking process.

One of the benefits of these stacked profiles is that they, more or less, represent the population of galaxies within their mass range; however, individual galaxies do deviate from the stacked construction.
Since individual profiles can show significant deviation in different locations based on a number of different factors, the stacked profiles should not be taken as representative of individual galaxies in Illustris or TNG.
We characterize individual galaxy profiles in Appendix \ref{appendix:idvGalaxies}.

\subsection{Fitting Methodology}
\label{subsec:Fitting}

The next step with these stacked median profiles is to characterize them with a fit. Typically, it is customary to employ a linear fit (linear in logarithmic metallicity space) within a specific region of a galaxy to calculate the metallicity gradient (observationally: \citeauthor{magrini_metallicity_2007} \citeyear{magrini_metallicity_2007}; \citeauthor{jones_measurement_2010} \citeyear{jones_measurement_2010}; \citeauthor{yuan_metallicity_2011} \citeyear{yuan_metallicity_2011}; \citeauthor{swinbank_properties_2012} \citeyear{swinbank_properties_2012}; and in simulations: \citeauthor{tissera_oxygen_2019} \citeyear{tissera_oxygen_2019}; \citetalias{hemler_gas-phase_2021}; \citeauthor{sharda_physics_2021} \citeyear{sharda_physics_2021}, \citeyear{sharda_origin_2021}; \citeauthor{yates_l-galaxies_2021} \citeyear{yates_l-galaxies_2021}).
However, the observations suggest that galaxies, beyond this region of a linear fit, flatten significantly \citep[][Q.-H. Chen et al., in prep, etc.]{martin_oxygen_1995,twarog_revised_1997,carney_elemental_2005,yong_elemental_2005,yong_elemental_2006,bresolin_flat_2009,vlajic_abundance_2009,vlajic_structure_2011,sanchez_characteristic_2014, grasha_metallicity_2022}.
Our goal in fitting these profiles is to, for the first time, characterize where this transition from a steep inner gradient to a weaker outer gradient occurs, which we designate the break radius (\rbreak).
The fitting method for our profiles, therefore, consists of a non-linear fitting routine described below.

We employ a Savitzky-Golay \citep[][]{savitzky_smoothing_1964,press_numerical_2007} smoothing to our median profiles.
The Savitzky-Golay Scipy cookbook\footnote{\href{https://scipy.github.io/old-wiki/pages/Cookbook/SavitzkyGolay}{https://scipy.github.io/old-wiki/pages/Cookbook/SavitzkyGolay}} method has three key parameters that could potentially influence the identified break radii in the galaxies.
The parameters input to this method are the order of the spline, the window size, and the smoothing factor.
Our nominal choices for these parameters for stacked profiles were to start smoothing over 9 points, corresponding to 0.9 kpc, with a seventh-order polynomial and smoothing factor of 0.1.
Individual profiles (see Appendix \ref{appendix:idvGalaxies}) deviate slightly from this in smoothing factor, which is increased to 0.25 as individual profiles have more noise-dominated features than the stacked profiles.
The starting window size of 0.9 kpc was chosen as it is \edit{larger, but not significantly so, than the the Illustris and TNG smoothing lengths (see Table~\ref{tab:resolutionTable})}.
These spline fits closely follow the median metallicity profiles in the presence of noise, allowing for cleaner calculations of the derivatives (gradients).
An example of this method applied to stacked profiles can be seen in the top panel of Figure \ref{fig:median_profiles}.
\edit{The bottom-left panel of Figure \ref{fig:OooohPretty} and the bottom panel of Figure \ref{fig:median_profiles} are examples of the individual galaxies methodology} (we again note, this analysis primarily focuses on results from stacked profiles, but individual profiles are discussed briefly in Appendix \ref{appendix:idvGalaxies}).
As can be seen \edit{from these examples}, the characteristic shape of a steep inner profile and shallow outer gradient holds for stacked profiles in a similar fashion to individual profiles.

From our fitted and smoothed profiles, we can compute the numerical gradient.
We define \rbreak{} of these stacked median profiles as the radius at which their gradients transition from steep to flat.
From the work of \citetalias{hemler_gas-phase_2021}, the inner gradient (defined by a linear regression over a central region of the star-forming gas in a similar sample of galaxies) was found to be well characterized by the following relation:

\begin{equation}
    \alpha = -\frac{C}{R_{\rm SFR}}.
    \label{eqn:ZachGradient}
\end{equation}

\noindent In this relation, $\alpha$ represents the gradient in dex/kpc, $R_{\rm SFR}$ is the radius enclosing 50\% of the star formation within a galaxy and $C$ was a free parameter equal to 0.28 across mass and redshift.
We define \rbreak{} as the location where the inner gradient (as identified in \citetalias{hemler_gas-phase_2021}) has decreased by a factor of 3.5.
We choose this factor of 3.5 as we expect the flat outer gradients to be roughly an order of magnitude shallower than the inner gradients, thus define the break as where the gradient is roughly half-way between the two, in log-space.
Changing this factor shifts the quantitative locations of the break radius, but does not significantly impact the qualitative trends observed in our analysis.
We can therefore define the value of the gradient at which the transition occurs as

\begin{equation}
    \color{\editcolor}\alpha_{\rm break} = -\frac{C_{\rm break}}{R_{\rm SFR}}%\left(\frac{R_{\rm SHM}}{R_{\rm SFR}}\right).
    \label{eqn:TNGgrad}
\end{equation}

\noindent We adopt a value of $C_{\rm break}=0.28/3.5=0.08$.
The ratio of the different characteristic lengths is computed for each stacked galaxy, using the median of the individual galaxies for each mass bin. 

As shown in Eqn.~\ref{eqn:ZachGradient}, inner gradients in TNG scale with $R_{\rm SFR}$, despite this, we utilize $R_{\rm SHM}$ as a proxy for galaxy size for a subtle reason.
In \citetalias{hemler_gas-phase_2021}, the gas samples are restricted to regions denser than the aforementioned star-forming density in TNG.
We do not employ this same cutoff and therefore chose a characteristic size not directly related to the star-formation of the galaxy.

% \subsubsection{Overfitting}
% \label{subsubsec:Overfitting}

% While developing our method of fitting the metallicity profiles with our combined Savitzky-Golay and spline fitting routine, we ran into the issue of overfitting our profiles.
% Looking further out from the centre of a galaxy, the metallicity begins to have significant dispersion (this can be seen in the 2D histogram on the bottom left panel of Figure \ref{fig:OooohPretty} and with the increased size of the error bars with radius in Figure \ref{fig:median_example}).
\edit{
Variation in these profiles exists on both small and large scales.
On the smallest scales the variation can arise from numerical effects or small structure deviations within the galaxy.
Though some of this variation is physical, this work's aim is to characterize the overall trend of metals in galaxies, not the detailed fine structure within the galaxies.
Therefore, in order to avoid sensitivity to these small-scale variations, we allow the Savitzky-Golay window size parameter to vary as we fit the profile.
The window size of the smoothing sets how many data points for the kernel to smooth over and is initially set at 0.9 kpc, corresponding to 9 data points. 
For each break radius determination, the window size is adjusted by $\pm 0.2$ and $\pm 0.4$ kpc (i.e. $\pm$ 2 and 4 data points) to produce additional fits for the profile.
From these four smoothed fits, as well as our original, we obtain five break radii for the same system and quantify how much they differ.
If these five fits generate break radii that vary significantly (nominally $>0.5$ kpc), then we assert that we are dominated by small scale noise of the system and reiterate the process by increasing the nominal window size by 0.2 kpc (2 data points) until convergence is attained.
Once agreement is found, we determine that we are no longer dominated by the small-scale structure and the mean of the five breaks is assigned the true {\rbreak} of the system.
If the window size starts to smooth over a significant portion of the galaxy (i.e. $\geq$10 kpc) and the identified break radii have not converged, we state that the system does not have a break radius.
We note that this maximum smoothing constraint does not impact the stacked median profiles, but does play a role in individual profiles (see Appendix \ref{appendix:idvGalaxies}).
}

% In order to avoid fitting more noise than data, we set constraints to appropriately handle the noise.
% These constraints are employed in the Savitzky-Golay smoothing routine.
% The window size of the smoothing, one of the free parameters of this routine \edit{(initially set at 0.9 kpc)}, is adjusted by \edit{$\pm 0.4$ and $\pm 0.2$ kpc to produce} different fits for the profiles.
% From these \edit{four perturbed} fits, as well as our original, we obtain a break radius from each and quantify how much they differ.
% If agreement is found (\edit{nominally 0.5 kpc agreement between the five fits}), we accept the mean as \edit{the true} {\rbreak} \edit{of the system}; however, if the values disagree significantly, we increase the \edit{base} smoothing length by 0.2 kpc \edit{until our answers converge}.
% \edit{In this process,} if the smoothing length gets too large (10 kpc) and the identified break radii have not converged, we state that the system does not have a break radius.
% We note that this maximum smoothing constraint does not impact the stacked median profiles, but does play a role in individual profiles (see Appendix \ref{appendix:idvGalaxies}).
% This process ensures that different evaluations of the noise do not artificially influence the location of {\rbreak}.

\subsection{Relevant Time-scales}
\label{subsec:Timescales}

% \sout{ In order to investigate the cause of the break, we examine how long the dominant effects in a system (mixing and enrichment) take.}
In order to investigate the cause of the break in the metallicity profile, we define the timescales for mixing and enrichment as a function of radius within our galaxy population.
We first create enrichment time-scale profiles for each stacked profile to measure how quickly star-formation is creating and ejecting metals into the ISM.
We define the enrichment time-scale, $\tau_Z$, as the fraction of current metallicity over the rate of change of the metallicity,

\begin{equation}
    \tau_Z = Z{\left(\cfrac{{\rm d}Z}{{\rm d}t}\right)}^{-1},
\end{equation}
where $Z$ is the gas-phase metallicity of the system.
To compute the rate of change of this metallicity $\rm{d}Z/\rm{d}t$ within the system, we use a closed-box model approximation (we briefly mention how this assumption could impact results in Section \ref{subsubsec:OtherSimulations}).
$\rm{d}Z/\rm{d}t$ is defined by the gas mass of the system $M_{\rm gas}$, an assumed yield value $y$, and the star formation rate ${{\rm d}M_*}/{{\rm d}t}$:

\begin{equation}
    \frac{{\rm d}Z}{{\rm d}t} \sim \frac{1}{M_{\rm gas}} y \frac{{\rm d}M_*}{{\rm d}t}. 
\end{equation}

\noindent The adopted yield value, $y = 0.05$, in our closed box model comes from \cite{torrey_evolution_2019}.
In that work, the authors derive the global yield value for metals in TNG at $z \approx 0$ and state that it may slightly differ at higher redshifts.
We make the simplifying assumption that this effective yield is constant with redshift throughout our analysis. \edit{We additionally assume that the yield is the same in the original Illustris simulation, as well.}

We define $\rm{d}Z/\rm{d}t$ in shells with radius 0.1 kpc, similar to how we define metallicity profiles, in order to investigate how quickly enrichment occurs as a function of radius.
To achieve this in a stacked profile, we take the median gas mass and the mean star-formation rate within each shell.

% \sout{The central idea behind investigating the enrichment time-scale profiles of our systems is to compare to another characteristic time-scale of the system.
% To this end,}
Additionally, we define the radial gas mixing time-scale, $\tau_{\rm GM}$, to represent the bulk exchange of material within a particular region of the galaxy.
Radial motion time-scale estimates for metallicity profiles have been used in the literature previously in \cite{roy_dispersal_1995} and \cite{martin_oxygen_1995}.
In those works, an estimate is made with a random walk statistical argument.
Given that we can track gas motions more precisely in simulations, we encode this information as the ratio of the radius, $R$, to the median radial velocity of the gas, $V_{\rm rad}(R)$, in the profile at that radius,

\begin{equation}
    \radialTimescale = \frac{R}{V_{\rm rad}(R)}
\end{equation}

Both the enrichment and gas mixing time-scales vary with radius and thus can be evaluated anywhere within the galaxy.
We evaluate both time-scales at \rbreak.
Enrichment of the gas in these systems, particularly through star formation, establishes the metallicity gradient; however, the radial exchange of materials redistributes the enriched gas, thus flattening the gradient.
Thus, these time-scales provide two useful proxies for these competing processes that drive both the formation and destruction of metallicity gradients.

\begin{figure*}
    \centering
    \includegraphics[width=\linewidth]{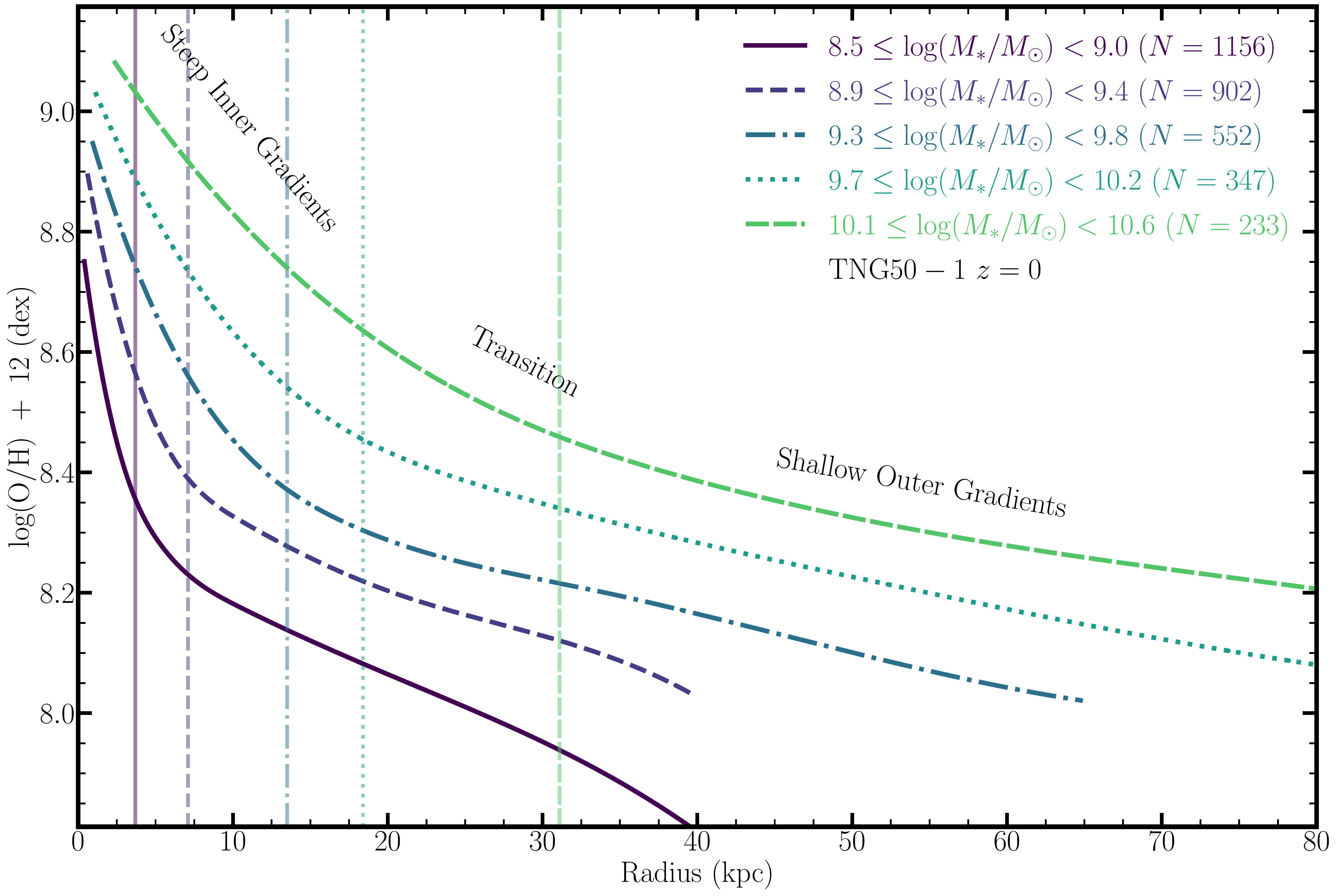}
    \caption{Five stacked median metallicity profiles at $z=0$ from the TNG50-1 star-forming galaxy population. Note that these stacked profiles are not all the same physical length, see Table \ref{tab:Simulated Distances} for how far each profile extends. Each profile demonstrates a steep gradient in the smallest galactocentric radii that flattens out at larger radii. Within the area of this transition between the two we define a break radius, \rbreak{} (Eqn. \ref{eqn:TNGgrad}). \rbreak{} for each profile is marked by the vertical line of the same color and linestyle of its corresponding profile.}
    \label{fig:LowZStacked}
\end{figure*}

\section{Results}
\label{sec:Results}

\subsection{Break Radii of Stacked Systems}
\label{subsec:BreakRadius}

\begin{figure*}
    \centering
    \includegraphics[width=\linewidth]{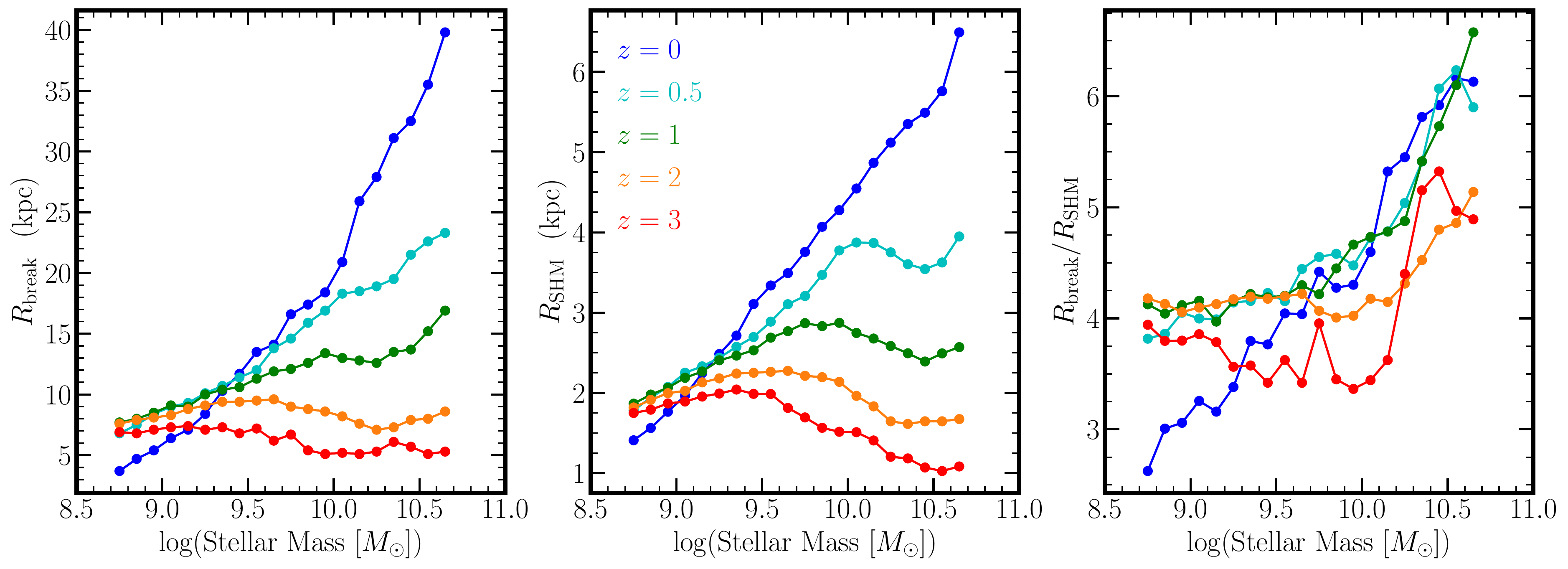}
    \caption{{\it Left:} The break radius as a function of mass and redshift for our sample of star-forming central galaxies in the TNG50-1 simulation (points plotted at geometric centres of bins). Each point represents the location at which the stacked metallicity gradient reaches the value set by Eqn. \ref{eqn:TNGgrad}. {\it Centre:} The median mass-size relation for galaxies as a function of mass and redshift \citep[stellar half mass radii were calculated in][]{genel_size_2018}. {\it Right:} The location of the break radius in each stacked profile, normalized by the stellar half mass radius.}
    \label{fig:TNG50_Relations}
\end{figure*}

Figure \ref{fig:LowZStacked} shows a sample of TNG50-1 $z=0$ stacked median profiles for a range of galaxy stellar mass bins.
We separate our galaxies into twenty different mass bins and generate a stacked median profiles for each bin (see Section \ref{subsec:ProfsGrads} for full details).
For each mass bin, we show the Savitzky-Golay smoothed spline fit as outlined in Section \ref{subsec:Fitting}.
We find that all of the stacked gas-phase metallicity profiles follow a trend of a steep inner gradient followed by a shallow outer gradient in TNG.
The vertical line marks the location of the identified break radius: the location where the gradient reaches the value set by Eqn. \ref{eqn:TNGgrad}.
Overall, our method accurately captures the radius at which the profiles transition from steep to shallow.
We find that the location of the break at $z = 0$ increases nearly monotonically spanning from $\sim5$ kpc at lower masses out past $\sim30$ kpc in the highest mass bins.

%Several previous works have suggested that metallicity gradient steepness has a relationship with galaxy stellar mass. 
%It follows from this that the location at which the gradient transitions from steep to shallow may also have some dependence upon mass. 
%To examine this relationship, 
%Figure \ref{fig:median_example} demonstrates the process of creating the stacked metallicity profiles for the $z=0$ $8.6\log M_\odot \leq M_* < 9.1\log M_\odot$ mass bin.
%The stacked median profile is in black with the shaded regions representing $1\sigma$ and 2$\sigma$ deviations with several randomly selected individual profiles overplotted in different colors.

We note that these stacked profiles are a construction used to analyse the bulk trends of galaxies within a certain mass range.
For example, most individual galaxies within our sample have this same trend of a weakening negative gradient with radius, but variations -- on large and small scales -- do exist.
Additionally, it has been seen in observational works that, at large radii, galaxies are also characterized by a region of near-constant metallicity \citep[][]{martin_oxygen_1995,twarog_revised_1997,carney_elemental_2005,yong_elemental_2005,yong_elemental_2006,bresolin_flat_2009,vlajic_abundance_2009,vlajic_structure_2011,sanchez_characteristic_2014}. We find that individual galaxies are well described by this flattening at large radii.
Figure \ref{fig:median_example} demonstrates a typical comparison between individual metallicity profiles and the resulting stacked metallicity profile for the $8.6 \leq \log(M_*/M_\odot) < 9.1$ mass bin at $z=0$.
The stacked profiles do not necessarily capture the flat outer metallicity profile that is frequently seen in individual galaxies as well as individual profiles; however the characteristic radius at which flattening occurs is similar for the stacked and individual profiles (see Appendix \ref{appendix:idvGalaxies}).
The lack of clarity capturing the flat outer metallicity profile is an expected feature of stacking the profiles, given that some galaxies within a mass bin exhibit a flattened gradient at lower radii than others. Beyond the flattened gradient the metallicity typically decreases again (Hemler et al., in prep), which explains the feature seen at the largest radii in the stacked profiles (see top panel of Figure \ref{fig:median_profiles}).
Combining these unique profiles means that such individualized features are smoothed out in the stacked profiles.

\subsection{Mass and Redshift Evolution}
\label{subsec:MassAndRedshiftEvo}

%Previous studies suggest that, in addition to having a dependence on mass, gradient steepness is also correlated with cosmic time (though whether increasing redshift weakens or strengthens the gradient is a point of contention).
%We again postulate that the gradient's relationship with redshift will lead to the break radius having the same, or similar, correlation with redshift.
%In order to examine this relationship, our sample of galaxies spans a large amount of cosmic time.
%We repeat the same analysis as Section \ref{subsec:BreakRadius} at $z = 0.5, 1, 2 ~{\rm and}~ 3$.

%Break radii were calculated using the same analysis as Section \ref{subsec:BreakRadius} at redshifts $z = 0.5, 1, 2 ~{\rm and}~ 3$.
The left panel of Figure \ref{fig:TNG50_Relations} shows the location of the identified break radius as a function of mass for galaxy populations at several redshifts.
At $z=0$ (blue line) we find that the break radius increases monotonically with increasing mass, which is consistent with the trend identified in Figure~\ref{fig:LowZStacked}.
The correlation between break radius and galaxy stellar mass holds going out to redshift $z=1$, but the steepness of the trend is significantly reduced; at $z>2$ there is virtually no mass dependence on the location of the break radius. 

% Given that each stacked profile within its given mass range is the same size (based on our definition of $r_{\rm max}$; Table \ref{tab:Simulated Distances}), a larger break radius indicates a weaker overall gradient (i.e. it takes the profile longer to flatten out).
% Thus, in general, the strength of the gradient within each profile increases with redshift, which qualitatively agrees with inside-out galaxy growth and previous theoretical studies of metallicity gradients in high-redshift galaxies (\citeauthor{pilkington_metallicity_2012} \citeyear{pilkington_metallicity_2012}; \citetalias{hemler_gas-phase_2021}; \citeauthor{yates_l-galaxies_2021} \citeyear{yates_l-galaxies_2021}).

%While the transition region, set by Eqn \ref{eqn:TNGgrad}, offers a fairly general prescription for the location of the break radius of stacked profiles, the actual transition occurs over a large region.
%Therefore, it is possible that the identification of break radii could be slightly different, yet still reasonable, with slightly different definitions. With that being said, the general trends observed in Figure \ref{fig:TNG50_Relations} are resilient to minor adjustments in the detailed location of the break radius.

%\subsection{Normalizing by a Characteristic Radius}
%\label{subsec:NormalizingBR}

Several works, both theoretical and observational, have noted that galaxy metallicity gradients appear to have weak mass dependence when normalized by some characteristic size of the galaxy (e.g. \citeauthor{sanchez_integral_2012} \citeyear{sanchez_integral_2012}, \citeyear{sanchez_characteristic_2014}; \citeauthor{sanchez-menguiano_shape_2016} \citeyear{sanchez-menguiano_shape_2016}; \citetalias{hemler_gas-phase_2021}).
Referred to as the `common abundance gradient', this relationship infers that the metallicity gradient is independent of the size of the galaxy.
It follows that the location where the profile flattens may, too, be independent of galaxy size.
We find that the characteristic radius that strongly correlates with the break radius is the stellar-half mass radius, \rshm.
{\rshm} is the 3D radius enclosing half of the stellar mass of the galaxy as in \citeauthor{genel_size_2018} (\citeyear{genel_size_2018}). 
The central panel of Figure \ref{fig:TNG50_Relations} demonstrates the mass and redshift evolution of \rshm{} in TNG50-1.
The stellar half mass radius increases nearly uniformly with galaxy stellar mass at low redshifts, yet at higher redshifts, this trend weakens significantly and turns over at the high mass end at $z>2$.
The reasons for the redshift evolution of galaxy size and eventual turnover at high redshift are beyond the scope of the paper; however, these trends closely follow those of mass and redshift identified for the break radius in Section \ref{subsec:MassAndRedshiftEvo}. 

The right panel of Figure~\ref{fig:TNG50_Relations} shows the break radius normalized by the stellar half mass radius. 
At redshift $z=0$ there is a remaining trend such that the normalized break radius continues to increase as a function of mass by a factor of $\sim$2 across our sampled mass range.
However, in the higher redshift bins, 
%When normalized by {\rshm} (see right panel of Figure \ref{fig:TNG50_Relations}), 
we find that the normalized break radius flattens as a function of galaxy stellar mass (i.e. break radius only weakly correlates with galaxy mass).
More specifically, by redshifts $z=0.5-1.0$ the normalized break radius is nearly flat over the mass range $8.5 < \mathrm{log} (M_* / M_\odot) < 10.0$, with a notable increase for the most massive galaxies.
For $z>1$, the normalized break radii are flat within a margin of $\pm25\%$ across the full mass range examined.
Critically, we note that the overall trend of the normalized break radius with mass is significantly weaker than was the case for the unnormalized break radius.
Whereas the break radius evolved by a factor of $\sim8$ at $z=0$, the normalized break radius evolves only by a factor of $\sim$2. 

In summary, we find that \rbreak{} increases linearly with mass at low redshifts and that this linear relationship flattens with increased redshift. When normalized by galaxy size, \rshm{}, the trends are significantly weaker.

%for nearly all redshifts it appears that for low- and intermediate-mass galaxies, there is no dependency on mass, but at higher masses the correlation appears roughly linear.
%The one apparent exception to this is $z=0$, where 

%There appears to be no clear correlation between the location of the break radius and redshift, however. 

\subsection{Time-scales at the Break Radius}
\label{subsec:BRTimescales}

\begin{figure}
    \centering
    \includegraphics[width=\linewidth]{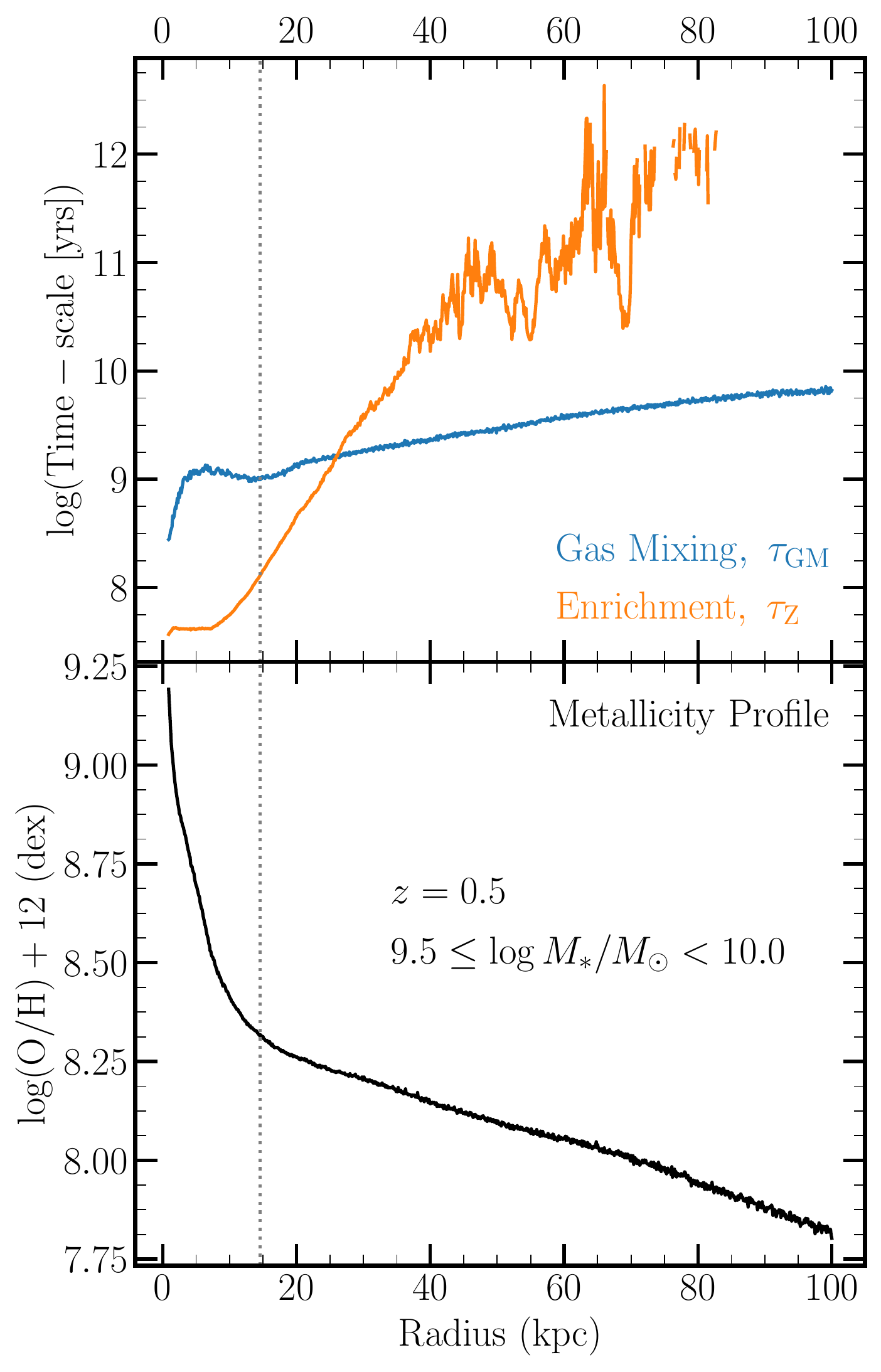}
    \caption{{\it Top:} Comparison of the gas mixing (blue) and enrichment (orange) time-scales (both defined in Section \ref{subsec:Timescales}) as a function of radius for the $9.5 \leq \log(M_*/M_\odot) < 10.0$ mass bin at $z=0.5$. The gaps in the enrichment time-scale profile correspond to radii containing negligible star-formation,  and therefore an infinite enrichment time-scale. {\it Bottom:} The corresponding gas-phase metallicity stacked profile for the same mass bin. From this profile, the break radius is computed (dotted gray line on both panels). We examine the ratio of the two time-scales at this location in Figure~\ref{fig:TNG50TimescaleComp}. }
    \label{fig:TimescaleExample}
\end{figure}

In order to understand the balance of physical processes giving rise to the break radius, we compare the enrichment and radial mixing time-scales of the gas within the system.
As stars form within a galaxy, they synthesize heavy elements which are eventually injected back into the ISM by stellar winds and SNe explosions.
This process drives up the local metallicity of the system and -- in the absence of mixing -- steepens the gradient.
Working in opposition to this, radial gas flows circulate the existing material throughout the galaxy.
Assuming roughly isotropic flows, metal-enriched gas is sent both towards the centre of the galaxy and out to the edges, effectively flattening the gradient.
The interplay of these processes dictates where the metals are and how they move through the galaxy.

\begin{figure}
    \centering
    \includegraphics[width=\linewidth]{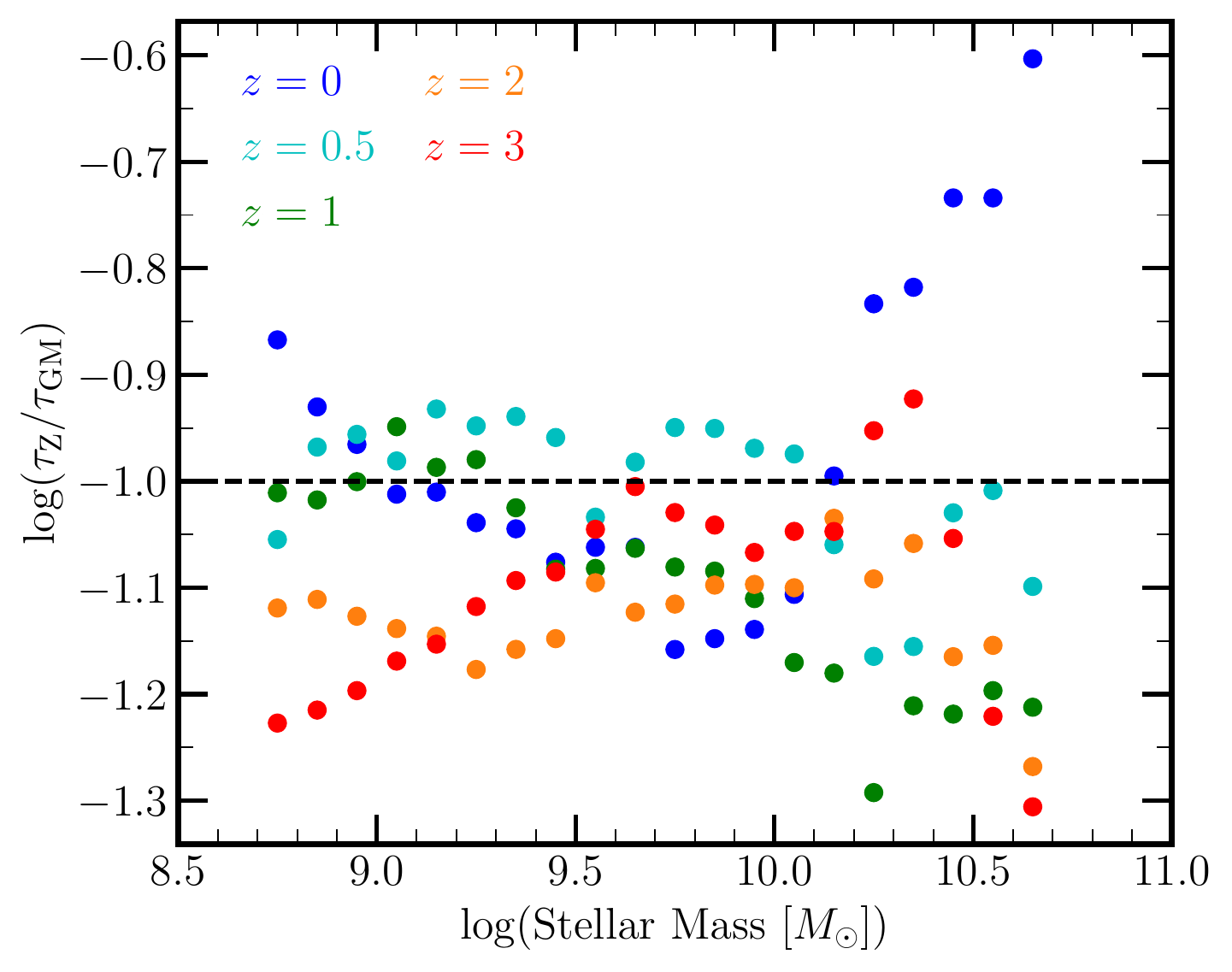}
    \caption{Ratio of enrichment time-scale $\tau_Z$ to radial gas mixing time-scales $\tau_{\rm GM}$ (both defined in Section \ref{subsec:Timescales}) at the break radius for stacked median profiles in the TNG50-1 varying with mass and redshift. The dashed line represents $\tau_{Z} = 0.1 \tau_{\rm GM}$.}
    \label{fig:TNG50TimescaleComp}
\end{figure}

For this analysis, we define two time-scales on which these two processes occur, the enrichment time-scale $\tau_Z$ and the radial gas-mixing time-scale $\tau_{\rm GM}$ (see Section \ref{subsec:Timescales}).
The top panel of Figure \ref{fig:TimescaleExample} shows these timescales as a function of radius for the $z=0.5$, $9.5 \leq \log(M_*/M_\odot) < 10.0 $ stacked profile. 
At small radii, $\tau_Z < \tau_{\rm GM}$, indicating that enrichment is far more efficient than the redistribution of the metals.
Traveling further out in the galaxy, the enrichment time-scale increases quickly, whereas the gas mixing time-scale increases much more slowly as a function of radius. 
This rapid increase in the enrichment time-scale can be attributed to the star-formation in the disk decreasing with radius within the disk.

We take particular interest in the ratio of these time-scales at the break radius (dotted gray line in Figure \ref{fig:TimescaleExample}; see Section \ref{subsec:Fitting}).
Figure \ref{fig:TNG50TimescaleComp} demonstrates this ratio as a function of mass and redshift.
We find that at the break radius, the ratio of these two time-scales is around $\log(\tau_{\rm Z}/\radialTimescale) = -1.0$, with scatter of approximately $\pm$0.3, for all mass and redshift bins.

The consistent ratio of enrichment to mixing time-scales at \rbreak{} hints that the location of the break in the metallicity profile may naturally arise from a competition between the mixing of the metals (which works to destroy gradients) and the enrichment via star-formation (which acts to build up gradients).
When one of these processes dominates over the other (i.e. is approximately an order of magnitude larger), the gradient follows accordingly.
Following from the definition of $\tau_{\rm Z}$ (and ${\rm d}Z/{\rm d}t$), in regions interior to break radius, as the star formation rate increases, the enrichment time-scale drops, and we find that enrichment easily dominates over mixing.  
In regions exterior to the break radius, the star formation rate drops and the enrichment time-scale becomes significantly subdominant compared against the mixing time-scale.  

\section{Discussion}
\label{sec:Discussion}

\begin{figure*}
    \centering
    \includegraphics[width=\linewidth]{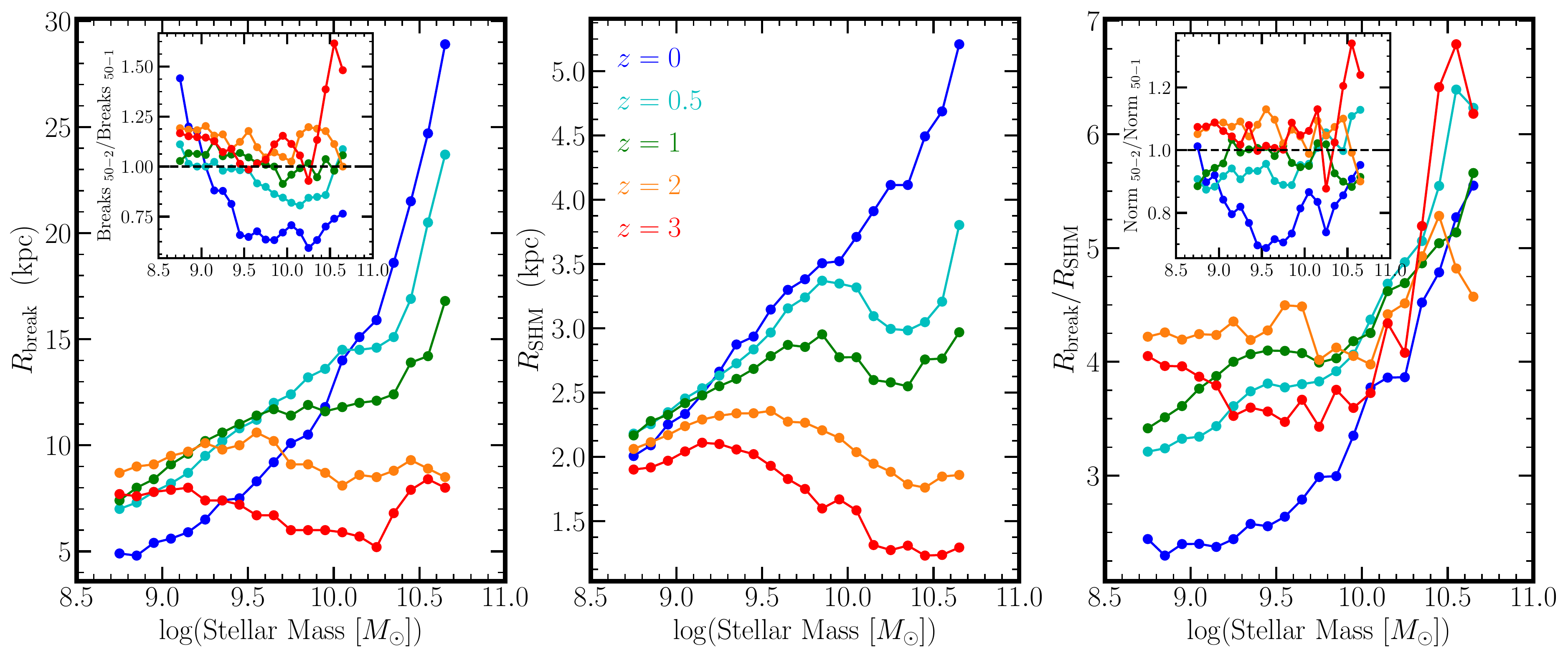}
    \caption{{\it Left:} The break radius as a function of mass and redshift for our sample of star-forming central galaxies in the TNG50-2 simulation (points plotted at geometric centres of bins). Each point represents the location at which the stacked metallicity gradient reaches the value set by Eqn. \ref{eqn:TNGgrad}. The inset represents a comparison of the location between the identified breaks in TNG50-2 versus TNG50-1, where the dashed black line shows complete agreement between the two simulations. {\it Centre:} The median mass-size relation for galaxies as a function of mass and redshift \citep[stellar half mass radii were calculated in][]{genel_size_2018}. {\it Right:} The location of the break radius in each stacked profile, normalized by the stellar half mass radius. Similar to the left panel, the inset on this panel represents the normalized break radii in TNG50-2 divided by the normalized break radii in TNG50-1.}
    \label{fig:TNG50-2_Relations}
\end{figure*}

\begin{figure}
    \centering
    \includegraphics[width=\linewidth]{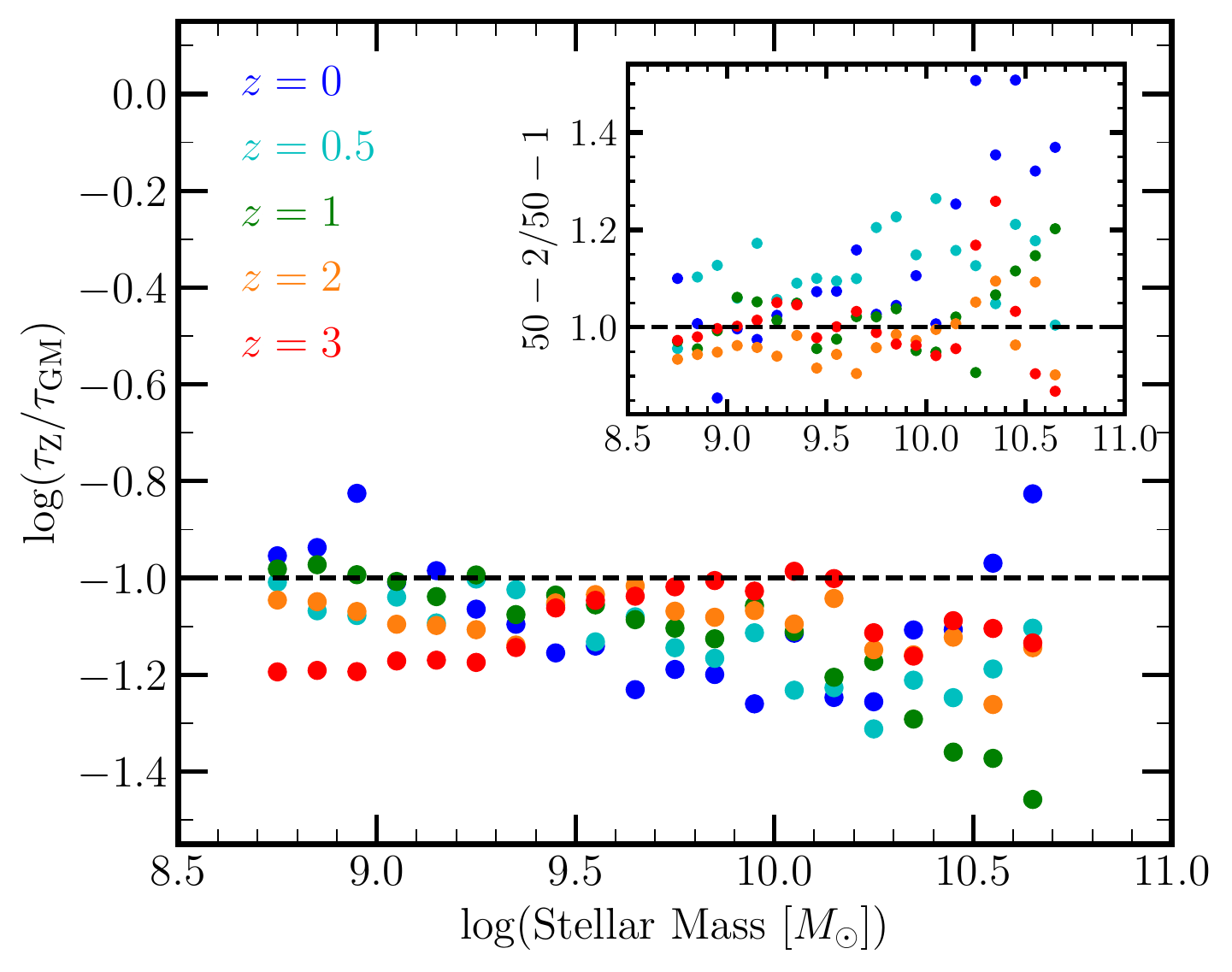}
    \caption{Ratio of enrichment time-scale to radial gas mixing time-scales (both defined in Section \ref{subsec:Timescales}) at the break radius for stacked median profiles in TNG50-2 varying with mass and redshift. The dashed line represents where the gas-mixing time-scale is an order of magnitude larger than the enrichment time-scale. The inset plot is a direct comparison to the results from the TNG50-1 time-scale analysis (see Figure \ref{fig:TNG50TimescaleComp}). The black dashed line on this plot shows where the two would agree completely.}
    \label{fig:TNG50-2TimescaleComp}
\end{figure}

\subsection{Dependence of Resolution and Adopted Physics}
\label{subsec:ResolutionAndAdoptedPhysics}

To investigate the role of resolution of the TNG50 simulation, as well as the physics in the model, upon our results, we repeated our analysis on TNG50-2, a lower resolution run of TNG50, and Illustris-1, the highest resolution run of Illustris, the predecessor to TNG.
The details of each of these runs are enumerated in Section \ref{subsec:Illustris}.
Notably, TNG50-1 and TNG50-2 are very similar runs, both with box sizes of (51.7 Mpc)$^3$, however, the number of resolution elements in TNG50-1 ($2\times2160^3$) is decreased to $1080^3$ in TNG50-2.
Based on our sample selections, at $z=0$ TNG50-2 has 2,083 galaxies while at $z=3$ it has 1,340 galaxies both of which are similar to the number of galaxies within the TNG50-1 simulation, as they are the same box size.
Furthermore, the breakdown of galaxies within each mass bin is also similar to that of TNG50-1.
We utilize this lower resolution run of TNG50 to determine whether resolution effects have any role in our findings. 
Similarly, Illustris-1 and TNG50-2 have a very similar resolution \edit{(see Table~\ref{tab:resolutionTable}, \citeauthor{vogelsberger_introducing_2014} \citeyear{vogelsberger_introducing_2014}, and \citeauthor{pillepich_first_2019} \citeyear{pillepich_first_2019} for more details)}, thus making them a relatively even-handed comparison.
However, the implemented galaxy formation physics models are sufficiently different from Illustris to TNG to allow us to probe the impact of the adopted galaxy formation model on the break radii properties.
We impose the same stellar mass and gas mass limits as the TNG50-1 sample, as well as bin these galaxies in the same manner.
Likewise, we impose the star-forming galaxy requirement as outlined in Section \ref{subsec:GalSelect}.
%Comparing the results of Illustris and TNG50 will allow us to understand the impact, if any, that the change in physics between the two models plays on setting the break radius of these systems. 
%The comparisons between TNG50-1 and TNG50-2, on the other hand, will allow us insight into whether or not the results are subject to the resolution of the simulations. 

%\subsubsection{Comparing Break Radii Across Simulations}
%\label{subsec:ComparingBRIllustrisAndTNG}

%We perform the same analysis for calculating the break radius as in TNG50-1 on the other two simulations of interest: TNG50-2 and Illustris-1.
%Similar to TNG50-1, we find that for TNG50-2 and Illustris-1 the vast majority of the galaxies within our sample have a steep inner gradient that smoothly transitions into an outer flat gradient. 
%We again take this region of transition to be the break radius of the system.
%These trends, shown by the individual galaxies, display in stacked profiles, as well.

\subsubsection{Resolution Effects} %TNG50-2
\label{subsubsec:Resolution}
Figure~\ref{fig:TNG50-2_Relations} shows the break radius versus mass, size-mass relation, and normalized break radius versus mass for the TNG50-2 simulation.
The results from Figure~\ref{fig:TNG50-2_Relations} can be compared against those presented in Figure~\ref{fig:TNG50_Relations} which shows the same results for TNG50-1 (explicitly shown in the inset plot in the left panel of Figure~\ref{fig:TNG50-2_Relations}).
%For the TNG50-1 and TNG50-2 samples (left panel of Figures \ref{fig:TNG50_Relations} and \ref{fig:TNG50-2_Relations}, respectively), 
We find excellent agreement in qualitative trends between physical-space break radius, in fact the majority of the break radii agree within $\sim25\%$.
We see that at $z=0$ the location of the break radius is directly correlated with mass, just as with TNG50-1, though the locations of the break radii vary the most at this lowest redshift.
We additionally see this relationship weaken with redshift and even invert at higher $z$.
The detailed locations of the break radii appear slightly larger (see inset in left panel) at lower masses, but (for all but $z=0$ and $3$) at higher mass the locations appear to match quite closely.
The $z=0$ break radii locations stray the furthest from the TNG50-1 sample

The galaxy sizes \citep[from][]{genel_size_2018} agree as well, in terms of qualitative trends, across the two simulations (central panels of Figures \ref{fig:TNG50_Relations} and \ref{fig:TNG50-2_Relations}).
The actual sizes of the galaxies are very slightly larger in TNG50-2, however.
We again normalize our break radii by {\rshm}, our chosen proxy for galaxy size (left panels of Figures \ref{fig:TNG50_Relations} and \ref{fig:TNG50-2_Relations}).
As can be seen in the inset plot in the right panel of Figure~\ref{fig:TNG50-2_Relations}, we find excellent agreement between the two samples, the locations all agree within 35\% (in fact, most agree within $\sim$10\%, particularly at higher redshifts).
% In both of these samples, we find that the normalized break has weak dependencies on both mass and redshift.
% Additionally, both of these relations show that the number of scalelengths we find this break at lies somewhere in the range of four to five \rshm.
We do note that there is more dispersion in the number of scale-lengths each break is at across redshift in this lower resolution run than in the higher resolution run.
This dispersion arises in the difference in the detailed locations of the breaks, outlined in the previous paragraph.
In general, we conclude that the location of the break radius does not appear to be a feature of the resolution of the TNG50-1 run, as we obtain similar results in TNG50-2.

The results of the time-scale analysis for TNG50-2 can be seen in Figure \ref{fig:TNG50-2TimescaleComp}. 
The agreement between TNG50-1 (Figure \ref{fig:TNG50TimescaleComp}) and TNG50-2 is remarkable.
The ratio of the time-scales at the break radius for the TNG50-2 sample deviate $\lesssim20\%$ (the higher redshifts are even closer than this) from the TNG50-1 sample for nearly all mass bins.
In both resolution runs we see that the ratio of the time-scales, at low redshift, is inversely correlated with mass (though high-mass $z=0$ stray from this trend in both).
This correlation weakens with redshift, not unlike that of the location of the break radius, until flattening out at $z\gtrsim2$.
The value at which the correlation flattens out to in TNG50-2 is around -1.00, the same value as in TNG50-1.
This suggests that our conclusion that the break radius is the location where the bulk radial gas flows dominate the gas enrichment is not a feature of the resolution of the TNG50-1 simulation run.

\subsubsection{Adopted Physics}
\label{subsubsec:AdoptedPhysics}

\begin{figure*}
    \centering
    \includegraphics[width=\linewidth]{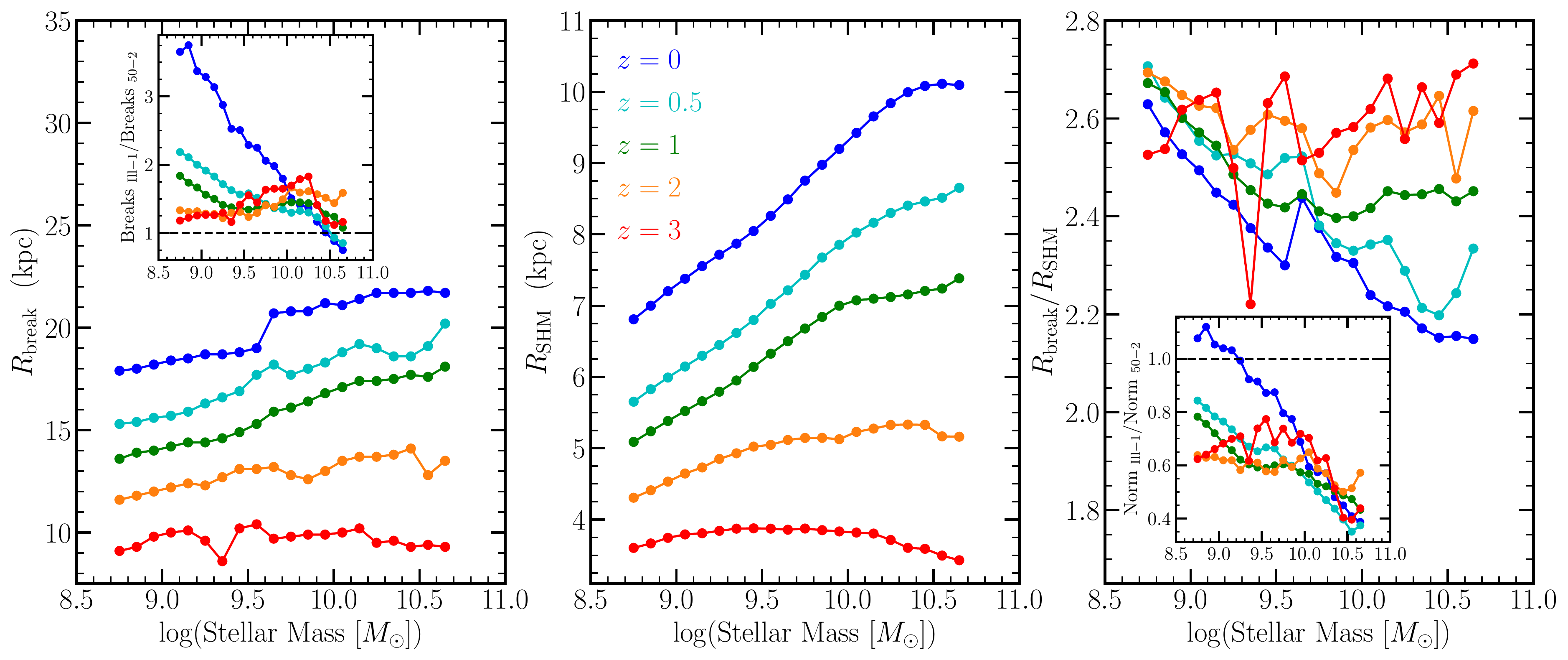}
    \caption{{\it Left:} The break radius as a function of mass and redshift for our sample of star-forming central galaxies in the Illustris-1 simulation (points plotted at geometric centres of bins). Each point represents the location at which the stacked metallicity gradient reaches the value set by Eqn. \ref{eqn:TNGgrad}. The inset represents a comparison of the location between the identified breaks in Illustris-1 versus TNG50-\edit{2}, where the dashed black line shows complete agreement between the two simulations. {\it Centre:} The median mass-size relation for galaxies as a function of mass and redshift. {\it Right:} The location of the break radius in each stacked profile, normalized by the stellar half mass radius. Similar to the left panel, the inset on this panel represents the normalized break radii in Illustris-1 divided by the normalized break radii in TNG50-1.}
    \label{fig:Illustris_Relations}
\end{figure*}

\begin{figure}
    \centering
    \includegraphics[width=\linewidth]{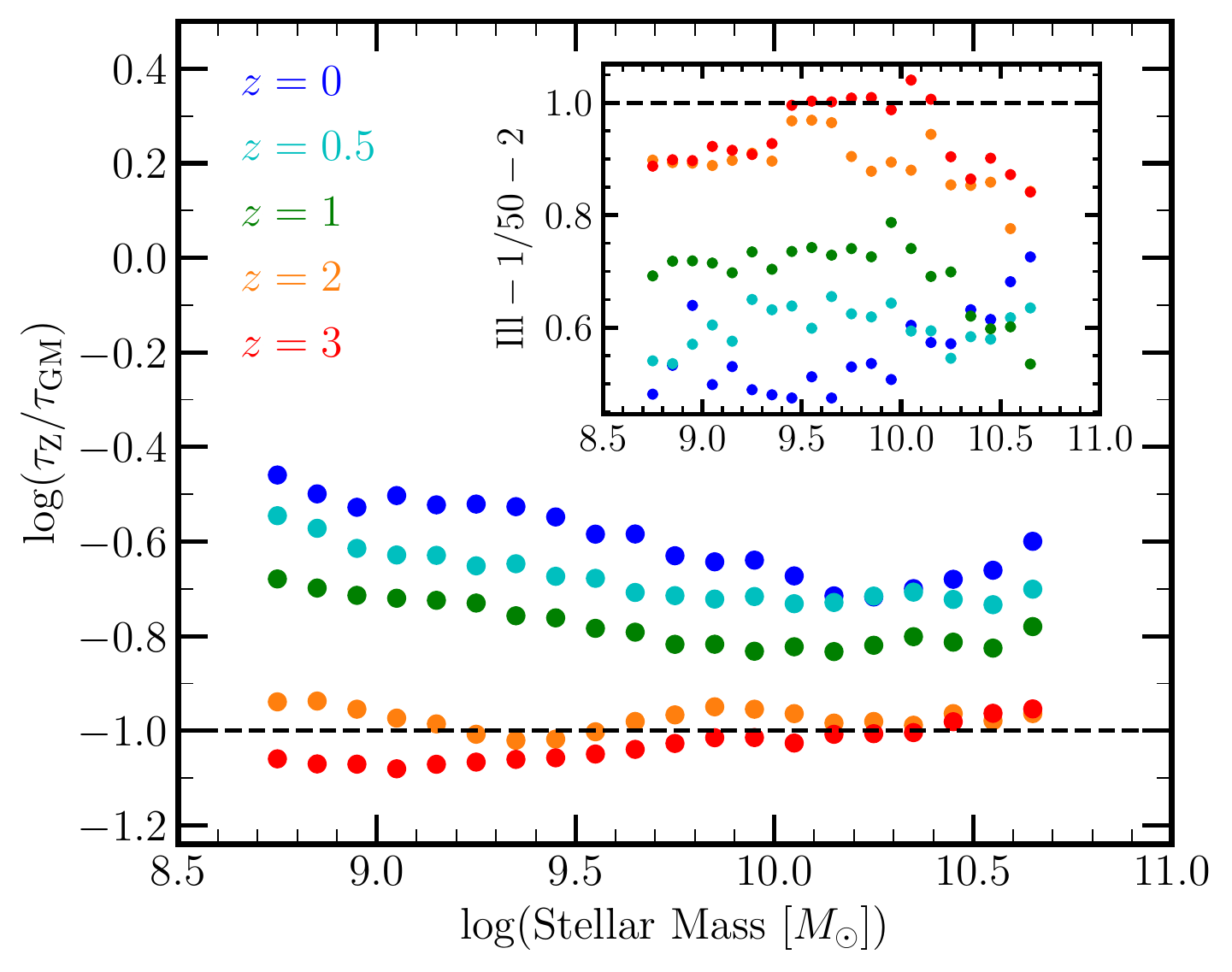}
    \caption{Ratio of enrichment time-scale to radial gas mixing time-scales (both defined in Section \ref{subsec:Timescales}) at the break radius for stacked median profiles in the Illustris-1 varying with mass and redshift. The dashed line represents where the gas-mixing time-scale is an order of magnitude larger than the enrichment time-scale. The inset plot is a direct comparison to the results from the TNG50-\edit{2} time-scale analysis (see Figure \ref{fig:TNG50-2TimescaleComp}). The black dashed line on this plot shows where the two would agree completely.}
    \label{fig:Illustris-1TimescaleComp}
\end{figure}

As a measure of the impact of the adopted physics of TNG, we use the original Illustris simulation suite as a comparison.
Of the varied physics in Illustris and TNG \citep[fully outlined in Section \ref{subsec:Illustris},][]{weinberger_simulating_2017,pillepich_simulating_2018}, the most important for our analyses are the changes to chemical enrichment as well as galactic winds.
% The main differences in the physical implementations of Illustris and TNG come from the galactic winds and stellar evolution (for full discussion of the difference of the physics between the two suites see Section \ref{subsubsec:AdoptedPhysics} as well as in \citeauthor{weinberger_simulating_2017} \citeyear{weinberger_simulating_2017} and \citeauthor{pillepich_simulating_2018} \citeyear{pillepich_first_2018}).
Additionally, Illustris has an increased box size (106.5 Mpc)$^3$ compared to TNG50 (51.7 Mpc)$^3$.
Based on our prescriptions, at $z=0$ Illustris-1 has 29,730 of galaxies and at $z=3$ it has 13,891 galaxies. 
The vast increase in galaxies compared to the TNG simulations is a feature of Illustris' larger volume.

When defining $r_{\rm max}$ (Section \ref{subsec:Radii}) we stated that the choice did not impact results as long as the transition region was captured within the region we examined.
This statement is important regarding the Illustris-1 $z=0$ sample. When following our prescription for $r_{\rm max}$ set by Table \ref{tab:Simulated Distances}, we do not fully capture this transition region.
For this reason, we extend all Illustris-1 $z=0$ galaxies out to 100 kpc.

We find that the break radii in the Illustris sample follow a weak positive relationship with mass (left panel of Figure \ref{fig:Illustris_Relations}), across redshift.
Additionally, the Illustris stacked profiles follow a much clearer redshift evolution and weaker mass dependence compared to the TNG samples.
We find the break radius decreases in all the mass bins across all redshifts.
Comparing this to the TNG50-\edit{2} sample (see inset on left panel of Figure~\ref{fig:Illustris_Relations}), we find that at higher redshift ($z\gtrsim$1), the location of the break radius is a factor of nearly 1.5 times larger across all mass bins. 
The lower redshifts show even more deviation from this, with $z=0$ varying from a factor of 5 times larger in the lowest mass bin, to half as large in the highest mass bins.

When normalizing the break radii by the characteristic size of the galaxy, \rshm, the residual mass and redshift trends are extremely weak (see the right panel of Figure \ref{fig:Illustris_Relations}).
We do note, however, that the number of scalelengths for which this break is identified is less than in TNG.
This is reflected in the inset for the right panel of Figure~\ref{fig:Illustris_Relations}, nearly all profiles' normalized break radii are smaller by $>20\%$.
The middle panels in Figures \ref{fig:TNG50_Relations}, \ref{fig:TNG50-2_Relations}, and \ref{fig:Illustris_Relations} show the evolution of the galaxy mass-size relation in each simulation.
It can be seen that the stellar half mass radius for the sample of Illustris galaxies is at least a factor of $\sim2$ larger than its TNG counterparts at all masses and redshifts.
Additionally, comparing the overall trend of the stellar half mass radii across simulations yields a similar result as with the break radii themselves: Illustris shows a much flatter relationship with a clear redshift evolution.
We, therefore, attribute some of this difference in galaxy size between the two simulations to cause the change in normalization.
This difference in galaxy size is one likely explanation for the deviation from the TNG50-\edit{2} sample. We conclude that the adopted physics do play a role insofar as setting the size of galaxies, which would, in turn, impact the location of the break radius.

The results from the Illustris-1 time-scale analysis can be seen in Figure \ref{fig:Illustris-1TimescaleComp}.
For each redshift, there is marginal correlation between mass and the ratio of the time-scales.
As a function of redshift, we find that the value of the flat correlation decreases with increasing redshift.
The highest redshifts, $z=2 ~{\rm and}~ 3$, correspond to the same time-scale ratio of the two TNG simulations, -1.00.
While the lower redshifts take a higher value for this time-scale, the values agree with low mass expectations from the TNG time-scales.
Additionally, these values of the time-scale ratios are all between $\sim -0.50 ~{\rm to}~ -1.00$ dex suggesting that the location of the break radius is where the radial mixing significantly dominates over the enrichment.

Despite the qualitative agreement, the time-scale ratios, particularly at low redshift, deviate roughly by a factor of two from the TNG50-1 sample (see inset panel of Figure~\ref{fig:Illustris-1TimescaleComp}). 
More specifcally, the Illustris sample shows this ratio to be less than in TNG50-1. 
This suggests that, while the location of the break radius is indeed set by the competition of mixing and enrichment, the exact relationship (i.e. an order of magnitude difference in time-scales) changes somewhat across physical models.

We, therefore, conclude that though a difference in physical models results in different locations of \rbreak{} and that the detailed interplay of processes is not perfectly consistent between models, the location of the break is determined by the opposition of mixing and enrichment within the disc.

% re are other factors (which are sensitive to the adopted physical model) that have a hand in setting the location of the break radius.

\subsubsection{Other Simulation Models}
\label{subsubsec:OtherSimulations}

This analysis centres entirely on the Illustris and TNG simulation suites; however, in principle, this analysis could be performed in any cosmological simulation presently available.
Both the Illustris and TNG models include a prescription for the star-forming regions of the ISM based on the \citeauthor{springel_cosmological_2003} (\citeyear{springel_cosmological_2003}, henceforth \citetalias{springel_cosmological_2003}) equation of state.
In this equation of state, a threshold density is set (the star-forming density $n_{\rm H} > 0.13$ cm$^{-3}$) above which the gas follows a set prescription for its pressure and temperature evolution. 
% (see Figure 1 of \citeauthor{torrey_evolution_2019} \citeyear{torrey_evolution_2019}).
In this way, all of the subgrid physics, the physics too small to resolve, is treated as the same.
Pressure support has the effect of keeping the gas from condensing and preventing further star-formation.
In the effective equation of state model, no distinctions are made (nor can be made) about the source of the pressure keeping the gas from condensing.
Therefore, different implementations of the ISM pressures could have a significant impact.

This exact treatment of the dense ISM is not implemented in all cosmological simulations and thus future analyses performed with different implementations could produce differing results.
For example, the EAGLE \citep[Evolution and Assembly of GaLaxies and their Environments,][]{schaye_eagle_2015,crain_eagle_2015} simulation suite follows the \cite{schaye_relation_2008} equation of state for the ISM.
This equation of state is fairly similar to the implementation in Illustris and TNG. Above a certain density, all gas falls onto a set curve on a density-temperature profile.
Since this equation of state resembles the \citetalias{springel_cosmological_2003} equation of state, we would only expect modest changes in the location of the break radius and our time-scale analysis.
We liken this to the analysis done here in Section \ref{subsubsec:AdoptedPhysics} comparing TNG and Illustris.
The locations of the break radius were not at the same physical location of the stacked galaxy profiles (even when comparing TNG50-1 to TNG50-2) nor were the exact trends that they each follow.
Yet we find that when the break radius is normalized, we see both simulations' relationships flatten out (albeit at a different value, though we attribute this to the size of the galaxies; see Section \ref{subsubsec:AdoptedPhysics}).

FIRE (Feedback In Realistic Environments, \citeauthor{hopkins_galaxies_2014} \citeyear{hopkins_galaxies_2014}), on the other hand, attempts to more explicitly model the multi-phase ISM.
The key to their approach is to (i) allow for gas to cool and condense into very dense gas clumps, (ii) restrict star-formation to be carried out within those dense gas clumps, and then (iii) implement stellar feedback locally around those young stellar populations.
In contrast to the \citetalias{springel_cosmological_2003} ISM treatment employed within Illustris and TNG, turbulence plays a significant role in the pressure support of the disk. 
This turbulence can act not only to support the gas disk against collapse, but can also radially redistribute metals and potentially wash out metallicity gradients.
The extent to which, e.g., bulk flows of gas versus turbulence help set the location of the break radius in simulated galaxies could provide an interesting -- potentially observable -- discriminator between models which significant turbulent ISM pressure support.

\edit{
Additionally, the FIRE model implements much burstier stellar feedback.
In these events, large bursts very quickly redistribute large amounts of gas throughout the disc, catastrophically destroying gradients.
However, since these events are so short-lived, the flattened gradient does not persist as the system re-equilibrates \citep[][]{ma_why_2017}.
This leads to a large diversity of gradients at high redshift ranging from flat to steep, which qualitatively agrees with observational studies \citep[e.g.][]{cresci_gas_2010,queyrel_massiv_2012,swinbank_properties_2012,wuyts_evolution_2016}.
Moreover, based on analytic arguments made in this paper, {\rbreak} should still be set by competition of enrichment and gas mixing within the disc, which would likely not significantly vary after the quick burst of outflowing material.
Thus, in a bursty feedback model, it is not clear that during the (re-)build-up of the gradient within the disc that the break radius should change significantly.

In comparison, non-bursty feedback models with discs dominated by bulk gas motions (i.e., Illustris and TNG) have no mechanism to rapidly destroy gradients.
Thus, as shown in \citetalias{hemler_gas-phase_2021}, these models show much steeper gradients at higher redshift compared to observations.
We therefore speculate that any change in the gradient would necessarily be set by a change in gas-mixing (e.g., mergers, change in winds, etc), which should have a direct effect on the location of the break radius.
We predict that the time variation of the location of the break radius (or the lack thereof) within a galaxy could also provide another constraint on simulated galaxy feedback models.
}

\subsection{Implications for the baryon cycle}

The baryon cycle is a central component in understanding galaxy evolution.
This process describes interactions between the ISM, CGM, and IGM, wherein gas accretes from the IGM and CGM onto the ISM and is expelled from the ISM via feedback.
% IGM gas cools onto the CGM and then onto the ISM.
% This gas, once accreted onto the ISM, provides fuel for star-formation within the disc. 
% These stars, in turn, enrich the gas and launch it back into the ISM and CGM via feedback (i.e. SNe, AGB winds).
Depending on the strength of the feedback, some of gas escapes into the IGM \citep[e.g.][]{heckman_absorption-line_2000} while some only makes it to the CGM and is `recycled' by cooling back into the ISM \citep[e.g.][]{shapiro_consequences_1976}.
Once back in the ISM, this process continues as the enriched gas mixes with new gas accreted onto the disc from the IGM. 

It is through this cycle that we present the ISM in two different parts: an inner star-formation dominated thin disc and an outer mixing dominated disc.
Star-formation in the inner disc leads to the production of metals.
Since star-formation occurs mostly in the centres of galaxies and less in the outskirts this metal production steepens the metallicity gradient.
% \citep[inside-out galaxy growth;][]{van_den_bosch_formation_1998,prantzos_chemo-spectrophotometric_2000,perez_evolution_2013},
On the other hand, the outer disc is subject to these ejected gases cooling and accreting onto it, as well as gas cooling from the IGM.
This region is therefore mixing dominated.
Additionally, since gas of all different stages of enrichment (recycled metal-rich gas from the inner ISM, metal-poor from the IGM) is being mixed into one place constantly, the gas is continuously being diluted, effectively flattening the gradient in this region.
The region where these competing effects meet is where the gradient transitions from steep to shallow -- i.e., the break radius.
Therefore, the break radius can be seen as one possible metric defining where the star-formation-dominated disc ends and the accreting, mixing-dominated outer disc starts. 

This follows from our time-scale analysis (Sections \ref{subsec:BRTimescales}, \ref{subsubsec:Resolution}, and \ref{subsubsec:AdoptedPhysics}), where we found, across nearly all mass, redshift, and in all examined simulations, that the location of the break radius is where the time-scale for gas enrichment is $\sim 1/10$ the time-scale for radial mixing.
Although we take a closed-box model approximation for our radial mixing time-scale, we would expect this statement to hold more generally as well.
In a more realistic system, gas is inflowing from the IGM and outflowing out of the CGM through winds and feedback.
These flows in and out of the galaxy would likely change the physical location of metals as well as contribute to the steepness of the gradient.
Barring any huge ejection or accretion events (i.e. mergers), the picture of break radius evolution should largely remain.
The enrichment from star-formation steepens the gradient in the inner portion of the disc, while the mixing flattens it.
Thus, the location where enrichment begins to dominate over mixing, regardless of model assumptions, should present a gradient transition region.

We discuss the formation model of the mixing dominated disc further in the companion paper by Hemler et al. (in preparation).

\subsection{How could break radii be observed?}
\label{subsec:Observed}

\subsubsection{Observational Challenges}

The convention for observationally determining metallicity gradients within a galaxy usually involves a linear-least squares regression though a limited region of a galaxy \citep[e.g.][]{magrini_metallicity_2007,jones_measurement_2010,yuan_metallicity_2011,swinbank_properties_2012,sanchez_characteristic_2014}.
This region, most typically [0.5, 2.0]$R_{\rm eff}$, falls almost entirely within the dense, star-forming regions of the ISM, by construction.
It is common for metallicity gradient surveys seek out \ion{H}{II} regions \citep[e.g.][]{shaver_galactic_1983,vilchez_chemical_1988,esteban_keck_2009,grasha_metallicity_2022} within the disc of the galaxy to measure nebular emission lines.
Therefore, observational studies with a focus on these regions only contain information on the star-forming inner disc of the galaxy.
Yet, as was shown in the previous sections (\ref{subsubsec:Resolution} and \ref{subsubsec:AdoptedPhysics}) as well as section \ref{subsec:BRTimescales}, the break radius is where star-formation becomes the sub-dominant process within the disc.
Thus, emission-line diagnostics could not provide enough spatial extent to robustly define a break radius for observed systems.

If we make a simplifying assumption that the stellar half mass radius is approximately equal to the effective radius\footnote{This assumption is not valid across all redshifts \citep[][]{suess_half-mass_2019}, but for the sake of simplicity we consider it to be the case.}, based on the right panels of Figures \ref{fig:TNG50_Relations}, \ref{fig:TNG50-2_Relations}, and \ref{fig:Illustris_Relations}, these observational studies would not encapsulate the break radius in any of the simulations we analyse. 
This, of course, is an exaggeration as the stacked profiles we present are just bulk trends and individual galaxies vary possibly having their break radius within $2R_{\rm eff}$ -- affording the ability to locate the break radius within these surveys. 
In fact, several observational studies have already noted the region where the metallicity flattens to a constant value \citep[][etc.]{martin_oxygen_1995,twarog_revised_1997,carney_elemental_2005,yong_elemental_2005,yong_elemental_2006,bresolin_flat_2009,vlajic_abundance_2009,vlajic_structure_2011,sanchez_characteristic_2014,grasha_metallicity_2022}.
Nevertheless, careful comparisons against observations are very difficult to make using emission diagnostics alone.

\subsubsection{Absorption Diagnostics}

An observational study designed to determine \rbreak{} would need to obtain metallicities from nearly the full extent of the galactic disc.
% If there were to be an observational study to determine a characteristic break radius, information from outside the typical star-forming region of the galaxy must be gathered.
% Such a study would need to create extended metallicity profiles.
One way to generate these profiles could be to combine emission diagnostics from the inner dense ISM with absorption measures in the extended, diffuse gas surrounding the disc. 
Absorption features can be observed in galaxies located at small angular separations from distant background quasars \citep[e.g.][]{werk_cos-halos_2012,werk_cos-halos_2014} in DLAs (Damped Lyman-$\alpha$ systems).
However, such systems are relatively rare. 
Even in systems with such background sources, the information is not very spatially extended, meaning that we are limited to only the regions directly in the line of sight. 
Stacking similar mass galaxies that absorption lines from background sources is one way to overcome this limitation \citep[e.g.][]{norris_oxygen_1983,ellison_enrichment_2000}.

With stacked profiles, an analogous set of analyses to those presented in this work could be performed on observational data. 
This, particularly with data from a number of different redshifts and stellar masses, would offer constraints on the current galaxy evolution models. 
For example, our new classification of inner star-forming discs and outer mixing dominated discs, mentioned in the previous section, could be verified.
Understanding whether observed galaxies follow this outlined prescription allows for yet another constraint to hold galaxies in cosmological simulations against.

\section{Conclusions}
\label{sec:Conclusion}

We select star-forming, central galaxies from the Illustris and TNG simulation suites with stellar mass limits of $8.5 < \log( M_*/M_\odot ) < 10.9$ broken into twenty mass bins.
We align these galaxies to a face-on orientation and create 1D gas-phase median metallicity profiles for each galaxy.
Individual galaxy profiles within each mass bin are then combined into a single stacked median profile.
From these stacked median profiles, we define the region where the gradient transitions from steep to shallow as the break radius. 
In order to understand where this break in the metallicity comes from we evaluate two relevant time-scales at the break radius: the enrichment time-scale and the radial gas mixing time-scale.

Our key results from this analysis are as follows:

\begin{itemize}
    \item We find that the majority of individual galaxy profiles exhibit a steep inner gradient followed by a shallow outer gradient. When the individual profiles are stacked, we see that this is the characteristic behavior of galaxies in our sample (see Figure \ref{fig:median_example}).
    \item Defining the region where this transition occurs as the break radius, \rbreak{}, we find that at $z=0$, \rbreak{} is positively correlated with the stellar mass of galaxies (Figure \ref{fig:LowZStacked}). This correlation weakens as a function of redshift, ultimately becoming negative at $z=3$ (left panel of Figure \ref{fig:TNG50_Relations}).
    \item The galaxy mass-size relation for the radius enclosing half of the stellar mass in the galaxy (\rshm) follows a similar trend as \rbreak{}. When the location of the break radius in the stacked profiles is normalized by {\rshm}, the mass and redshift dependencies flatten (right panel of Figure \ref{fig:TNG50_Relations}).
    \item In order to understand the interplay of physics flattening metallicity gradients, we define two relevant time-scales: a radial gas mixing time-scale ($\radialTimescale$) and a gas enrichment time-scale ($\tau_{Z}$). The ratio of these time-scales at the break radius yields a roughly constant value (with some weak residual trends) with both mass and redshift. This suggests that the flattening of the metallicity occurs when mixing becomes appreciably dominant compared to gas enrichment (Figure \ref{fig:TNG50TimescaleComp}).
    \item In TNG50-2, a lower resolution run of TNG50, we find good agreement between both the location of the break radius across mass and redshift as well as the ratio of time-scales, suggesting that our results are not features of the resolution of TNG50-1 (see Sections \ref{subsubsec:Resolution} and Figures \ref{fig:TNG50-2_Relations} and \ref{fig:TNG50-2TimescaleComp}).
    \item In Illustris-1, a different, but similar, physical model to TNG, we find the locations of break radii, normalized by galaxy size, are a factor of two smaller than TNG galaxies, though this can be explained by Illustris galaxies being approximately a factor of two larger. Additionally, we find that the ratio of time-scales generally agrees with our TNG50-1 analysis, though there is some deviation (see Section \ref{subsubsec:AdoptedPhysics} and Figures \ref{fig:Illustris_Relations} and \ref{fig:Illustris-1TimescaleComp}).
\end{itemize}

The general agreement between TNG50-1, TNG50-2, and Illustris suggest that the location of the break radius is indeed set by the competition between enrichment and mixing.
Observing these extended profiles, which would require the combination of absorption and emission diagnostics (see Section \ref{subsec:Observed}), could provide a potential discriminator for models of ISM pressure support in the future.

\section*{Acknowledgements}

% {\red Please add acknowledgement statements as necessary :)}
PT acknowledges support from NSF grant AST-1909933, AST-2008490, and NASA ATP Grant 80NSSC20K0502.
K.G. is supported by the Australian Research Council through the Discovery Early Career Researcher Award (DECRA) Fellowship DE220100766 funded by the Australian Government. 
Part of this work is supported by the Australian Research Council Centre of Excellence for All Sky Astrophysics in 3 Dimensions (ASTRO~3D), through project number CE170100013. 
% The Acknowledgements section is not numbered. Here you can thank helpful
% colleagues, acknowledge funding agencies, telescopes and facilities used etc.

%%%%%%%%%%%%%%%%%%%%%%%%%%%%%%%%%%%%%%%%%%%%%%%%%%
\section*{Data Availability}

The data that support the findings of this study are available upon request from the corresponding author. Most of the data from the Illustris and IllustrisTNG is already available on each project's respective website. Illustris: \href{https://www.illustris-project.org/data/}{https://www.illustris-project.org/data/} and IllustrisTNG: \href{https://www.tng-project.org/data/}{https://www.tng-project.org/data/}



\bibliographystyle{mnras}
\bibliography{references} % if your bibtex file is called example.bib

%\begin{thebibliography}{99}
%\end{thebibliography}

%%%%%%%%%%%%%%%%%%%%%%%%%%%%%%%%%%%%%%%%%%%%%%%%%%

%%%%%%%%%%%%%%%%% APPENDICES %%%%%%%%%%%%%%%%%%%%%

\appendix

\section{Individual Galaxies}
\label{appendix:idvGalaxies}

Though our analysis centers on the identification of break radii for stacked profiles as a means of potential comparisons with observations, individual galaxies can undergo this same analysis.
We noted in Section \ref{subsec:BreakRadius} that stacked and individual profiles do have their differences. 
In this appendix, we take a cursory glance at population trends for individual galaxies and compare how they hold up against our stacked profile analysis. For simplicity, we examine only the TNG50-1 sample in this work.

In Figures \ref{fig:OooohPretty} and \ref{fig:median_profiles}, we demonstrated fitting an individual profile.
The fitting methodology is nearly identical to that outlined in Section \ref{subsec:Fitting}, however minor changes were implemented to optimize the fitting of individual profiles. 
The first of such changes is that instead of using the value of $R_{\rm SFR}$ for the stacked profiles, we compute the gradient threshold, $\alpha_{\rm break}$ (see Equation \ref{eqn:TNGgrad}), using the individual galaxy's characteristic sizes. 
This avoids the ambiguity of overlapping mass ranges created when binning the profiles.
Additionally, since individual profiles are not as well behaved as stacked profiles (e.g. noise due to environment, etc.), we slightly modify our Savitzky-Golay smoothing to avoid overfitting to sharp features.
We therefore adjust the smoothing parameter from 0.1 to 0.25 for individual profiles.
Finally, as was the case for stacked systems, our method of computing the break radius is iterative in selecting a window size (see Section \ref{subsec:Fitting}), if this process starts to smooth over a significant portion of the galaxy -- 10 kpc -- and has not identified a clear break radius, we choose to stop the process and assign no break radius.
No stacked profile failed to identify a break radius, but we find that some individual systems did not have an identified break.
At the highest redshift ($z= 1,2, ~{\rm and}~3$), less than one percent of the total galaxies in our sample have no break radius identified. At $z=0.5$, 2.8\% of galaxies do not identify a break radius while at $z=0$, 8.6\% of galaxies are impacted.

\begin{figure*}
    \centering
    \includegraphics[width=0.45\linewidth]{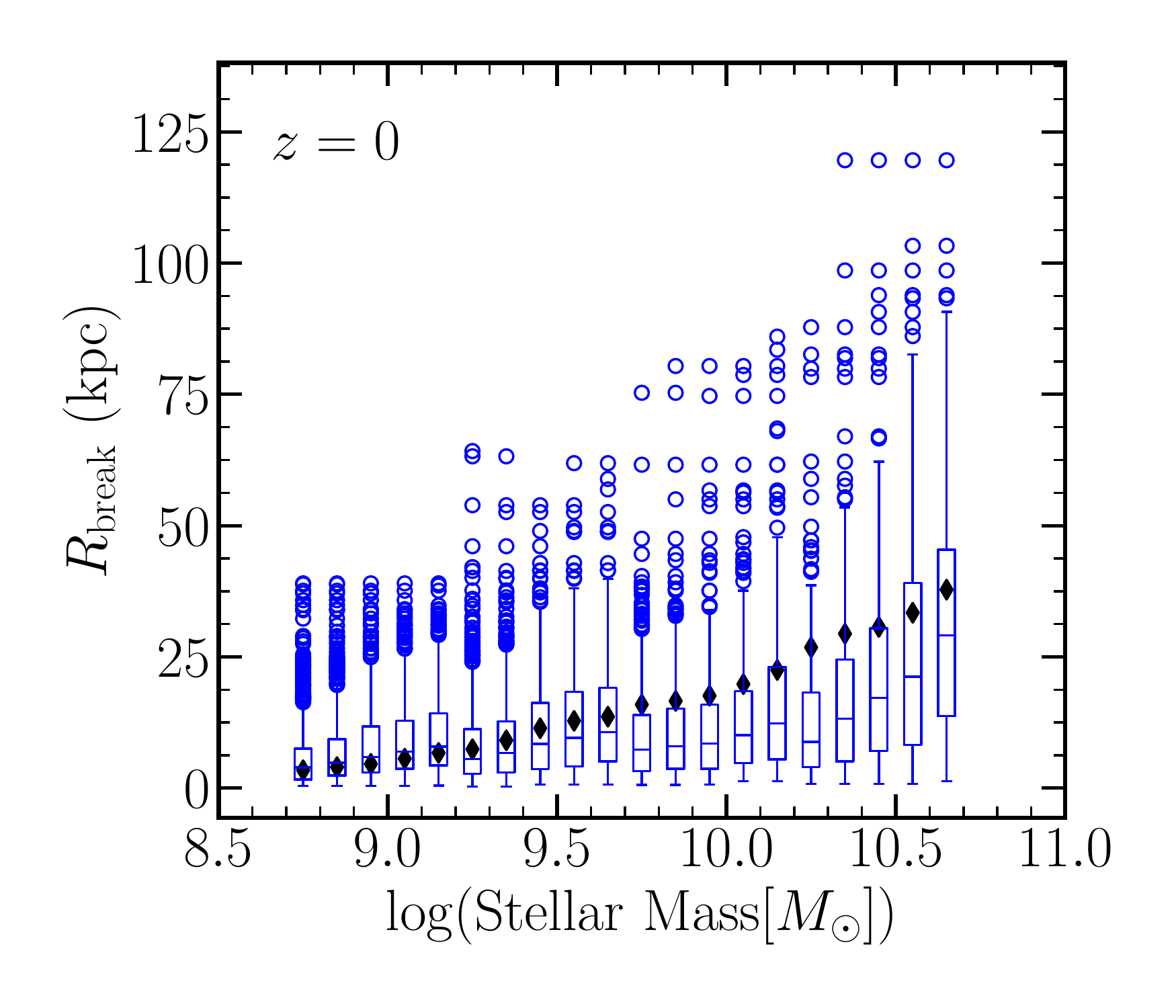}
    \includegraphics[width=0.45\linewidth]{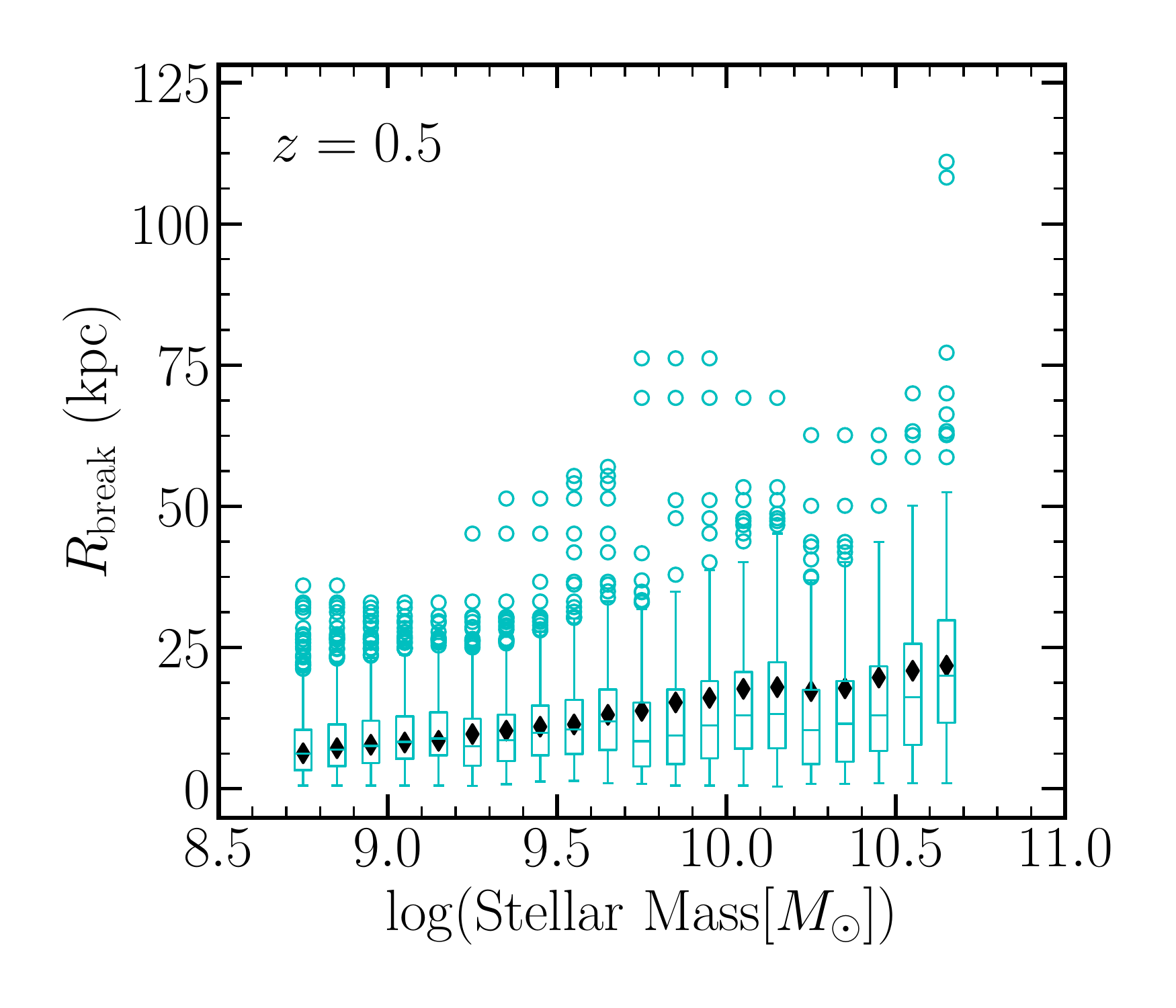}
    \includegraphics[width=0.45\linewidth]{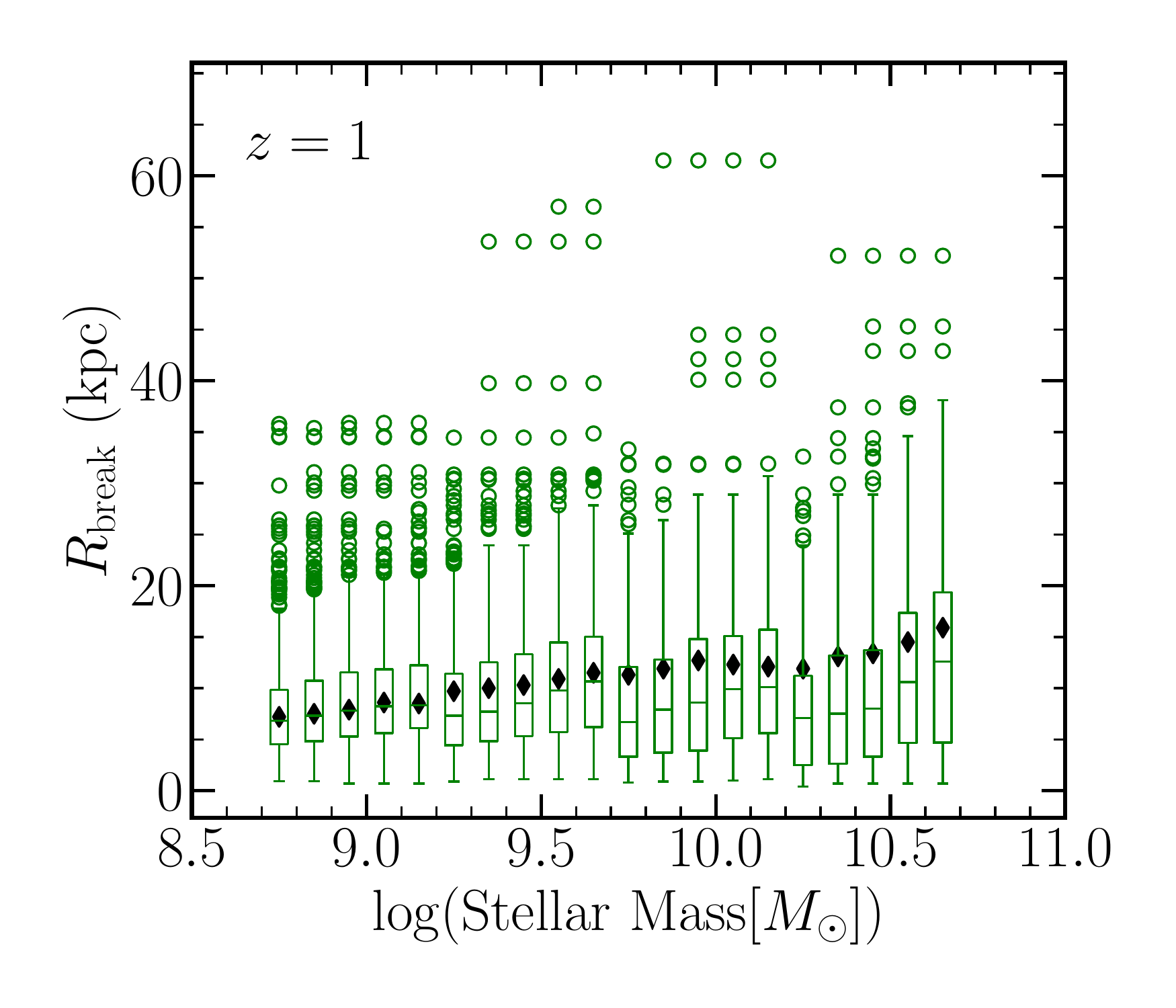}
    \includegraphics[width=0.45\linewidth]{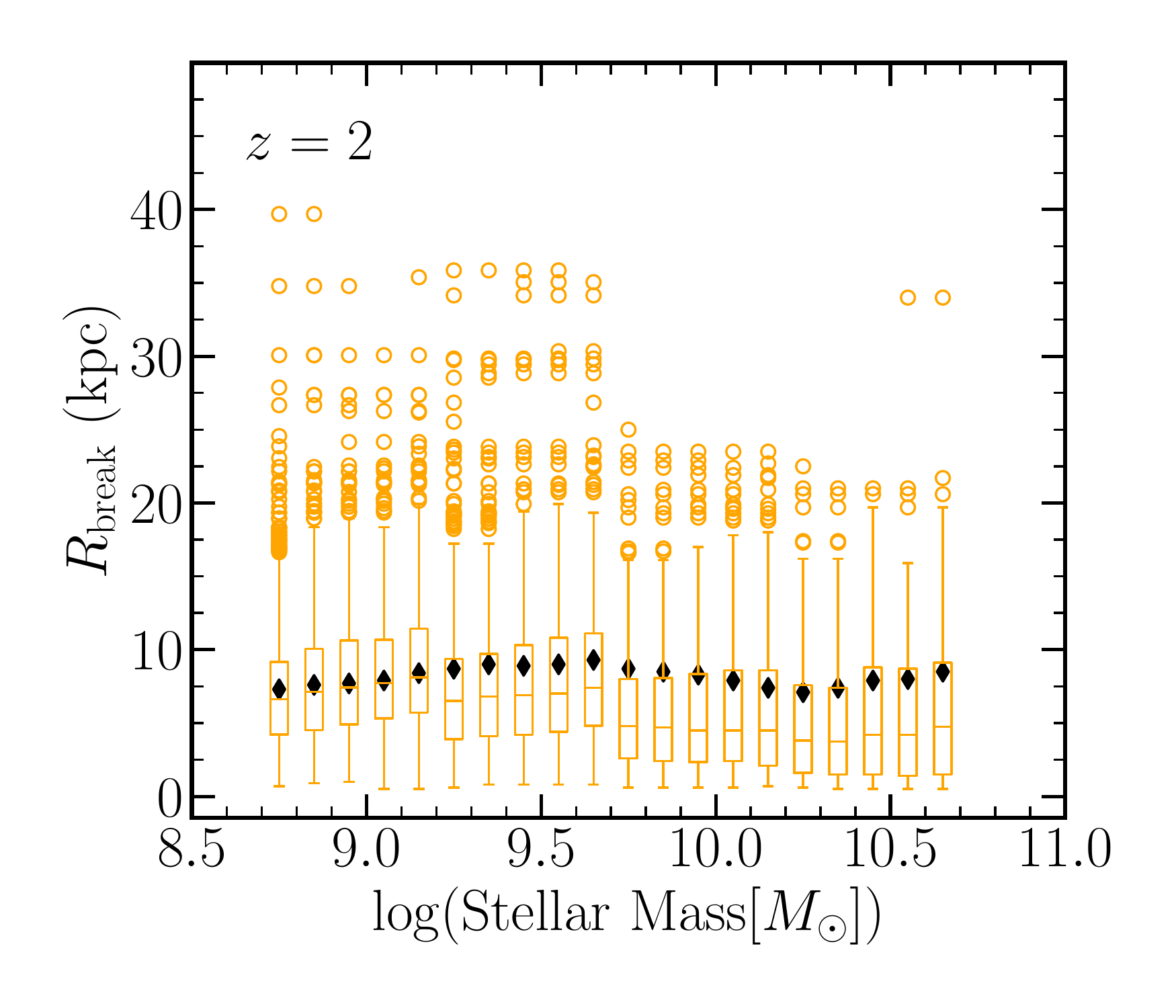}
    \includegraphics[width=0.45\linewidth]{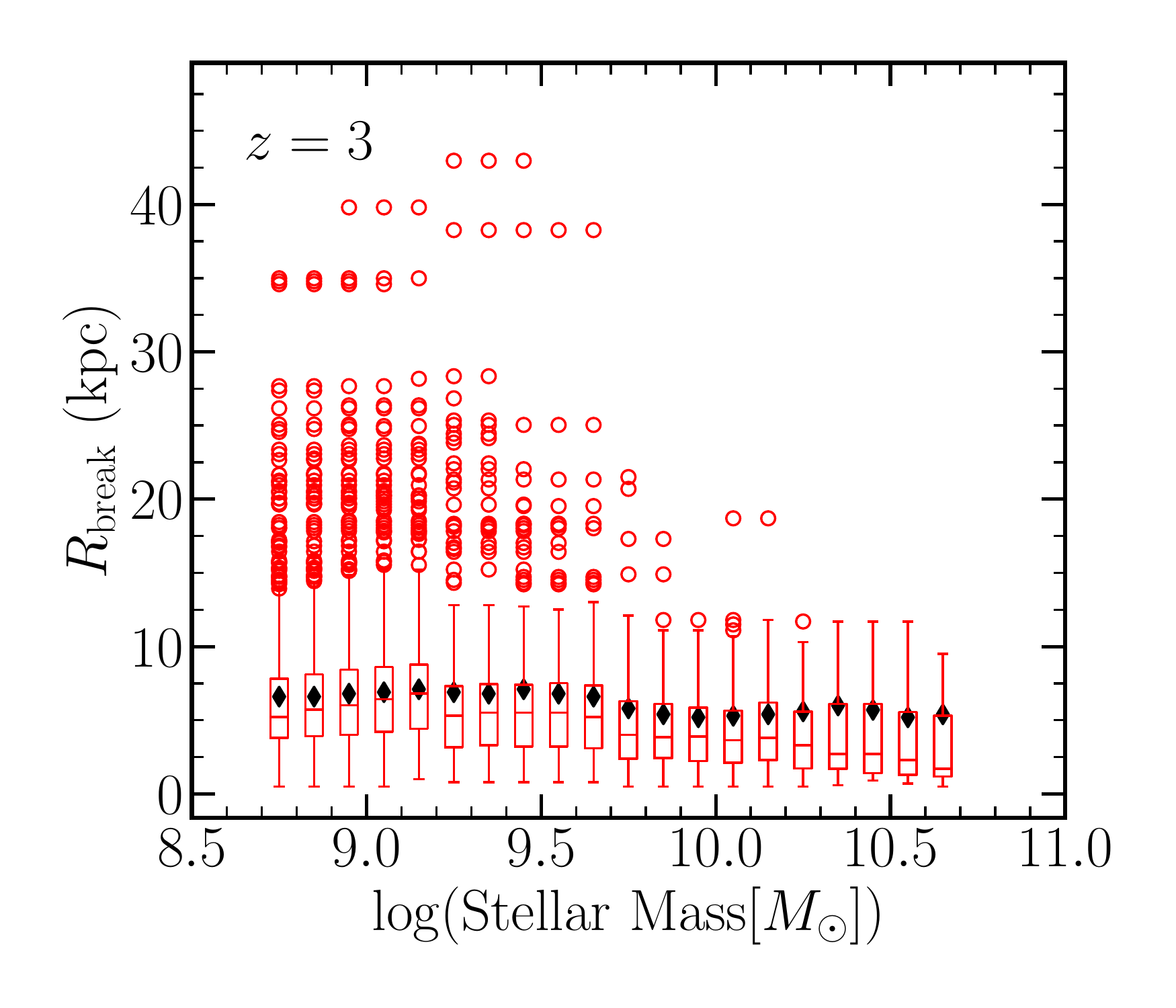}

    \caption{Box-plots of individual galaxies' \rbreak{} at all analyzed redshifts for the TNG50-1 star-forming galaxies. Each box is plotted in the geometric center of the mass bin it represents. The box represents the first and third quartiles, with the median plotted across it. The bars extending off the box are within 1.5$\times$ the two quartiles and the open dots represent all data outside that. The black diamonds represent the stacked profile's identified break radius (see section \ref{subsec:BreakRadius}).}
    \label{fig:IdvBreaks}
\end{figure*}

We find that for nearly all mass bins across all redshifts, the break radius for the stacked profiles generally falls in between the median and third quartile of individual profile's break radii (see Figure \ref{fig:IdvBreaks}).
\rbreak{} being systematically larger in stacked profiles is an expected behavior of stacking gas-phase metallicity profiles.
As was briefly discussed in section \ref{subsec:BreakRadius}, individual profiles all flatten out at different radii, thus by stacking them we expect that the median profile will be pushed further out by profiles with very flat gradients.

\edit{
We caution that while this method for finding the break radius of individual galaxies works reasonably well for most galaxies within our sample, there is variability.
We have inspected a number of samples by eye and find reasonable agreement with expectations.
}

\edit{
\section{Differences in stacking techniques with observations}
\label{appendix:stacking}

As mentioned in Section~\ref{subsec:ProfsGrads}, when generating stacked median profiles we first create median profiles for each individual galaxy then take a median-of-medians approach to get our result. 
% The reason we employ this method is due to the structure of the Illustris and IllustrisTNG datasets.
% It is more efficient to load in a galaxy and then work with that galaxy's data saving only the median profile.
% Though more useful in practice,
This approach does differ from the traditional methods of stacking in observational studies, wherein the metallicity profile (whether a median or weighted-mean) is calculated using all of the individual points \citep[e.g.,][]{leccardi_ICM_2008,carton_stacking_2015}.
While these two methods are indeed different, upon generating stacked profiles both ways in TNG50-1 at $z=0$, we find that the location of the break radius using either method agrees within $\sim20\%$ (see Figure~\ref{fig:breaksComp}).

\begin{figure*}
    \centering
    \color{\editcolor}
    \includegraphics[width=0.7\linewidth]{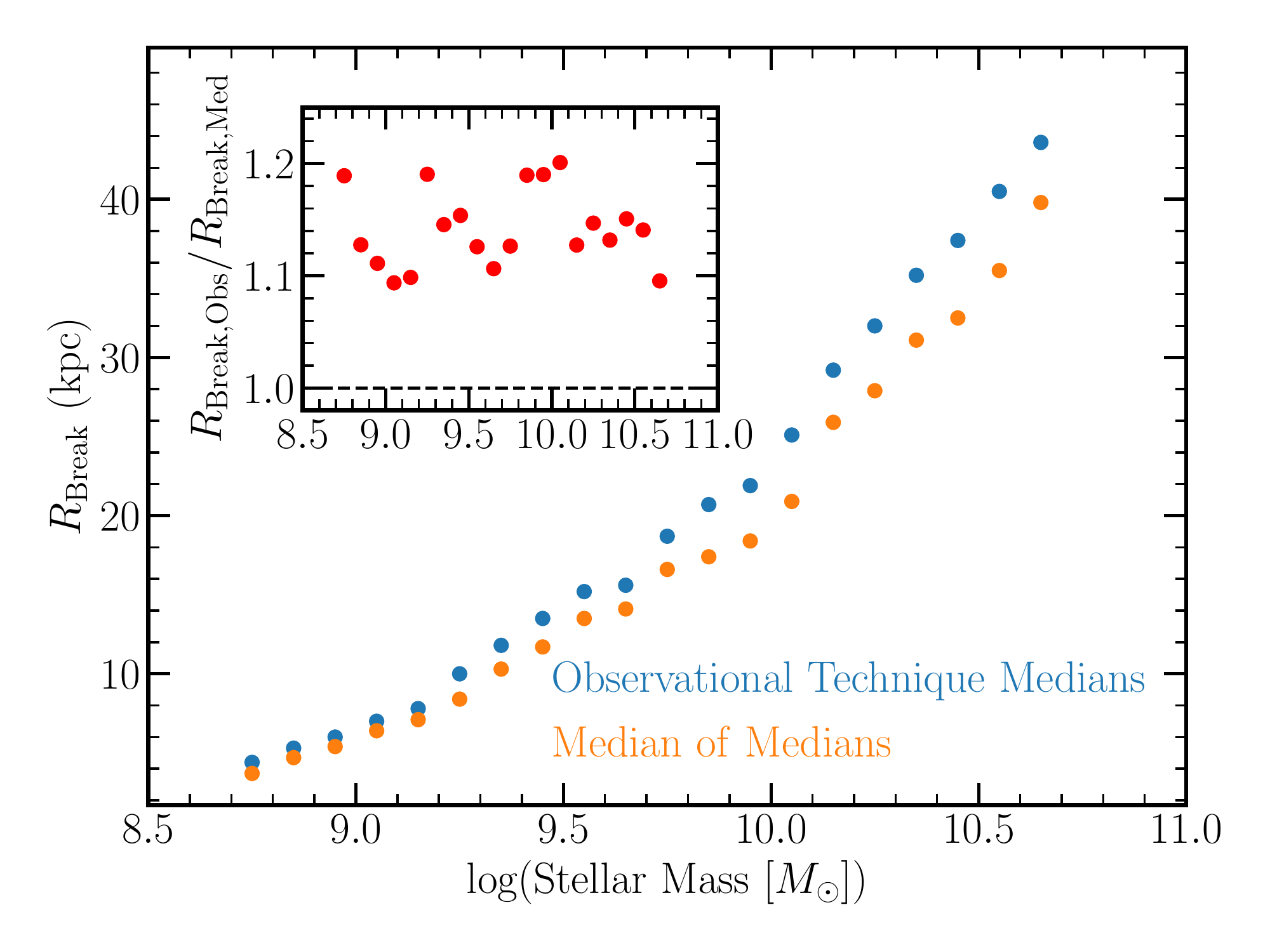}
    \caption{A comparison of the identified break radii using observational stacking techniques (wherein stacked median profiles are generated using all of the data from all of the galaxies; shown here in blue) against the technique of stacking individual median profiles (see Section~\ref{subsec:ProfsGrads}; shown here in orange) for each stellar mass bin in TNG50-1 and $z=0$. The inset shows the ratio of the observational approach over ours for each mass bin. }
    \label{fig:breaksComp}
\end{figure*}
}
% \section{Some extra material}

% If you want to present additional material which would interrupt the flow of the main paper,
% it can be placed in an Appendix which appears after the list of references.

%%%%%%%%%%%%%%%%%%%%%%%%%%%%%%%%%%%%%%%%%%%%%%%%%%

% Don't change these lines
\bsp	% typesetting comment
\label{lastpage}
\end{document}